\documentclass[journal,10pt,comsoc]{IEEEtran}
\usepackage[T1]{fontenc}
\usepackage{psfrag,amsmath,amssymb,color,cite, amsmath,graphicx}
\usepackage{amsthm}
\usepackage{listings}
\usepackage{stfloats}
\usepackage{mathtools}

\DeclarePairedDelimiter\floor{\lfloor}{\rfloor}

\newcommand{\DeclareAutoPairedDelimiter}[3]{%
  \expandafter\DeclarePairedDelimiter\csname Auto\string#1\endcsname{#2}{#3}%
  \begingroup\edef\x{\endgroup
    \noexpand\DeclareRobustCommand{\noexpand#1}{%
      \expandafter\noexpand\csname Auto\string#1\endcsname*}}%
  \x}
\DeclareAutoPairedDelimiter\modulo{[}{]} 
\usepackage{fancyhdr,graphicx,amsmath,amssymb}
\usepackage[linesnumbered,ruled,vlined]{algorithm2e}
\include{pythonlisting}
\usepackage{mathrsfs}
\usepackage{multirow}
\usepackage{pifont}
\usepackage[colorinlistoftodos]{todonotes}

\newtheorem{lemma}{Lemma}

\newcommand{\beqn}{\begin{equation}}
\newcommand{\eeqn}{\end{equation}}
\newcommand{\beqa}{\begin{eqnarray}}
\newcommand{\eeqa}{\end{eqnarray}}
\newcommand{\beqas}{\begin{eqnarray*}}
\newcommand{\eeqas}{\end{eqnarray*}}

\presetkeys{todonotes}{color=green!30}{}
\usepackage{amsmath}
\usepackage{amsfonts}
\usepackage{graphicx}
\usepackage{array}
\usepackage{ amssymb }
\newcolumntype{P}[1]{>{\centering\arraybackslash}p{#1}}
\newcolumntype{M}[1]{>{\centering\arraybackslash}m{#1}}

\newcommand{\thickbar}[1]{\mathbf{\bar{\text{$#1$}}}}
\newcommand{\thicktilde}[1]{\mathbf{\tilde{\text{$#1$}}}}
\begin{document}

\title{Low Complexity Iterative Rake Decision Feedback Equalizer for Zero-Padded OTFS systems}
\author{\IEEEauthorblockN{Tharaj Thaj and Emanuele Viterbo 
\thanks{A preliminary version of this work will be presented in part at the {\em IEEE Wireless Communications and Networking Conference (WCNC)}, May 2020, \cite{WCNC_paper}.
The authors are with Department of Electrical and Computer Systems Engineering, Monash University, Clayton, VIC 3800, Australia. E-mail: tharaj.thaj@monash.edu, emanuele.viterbo@monash.edu. This work was supported by the Australian Research Council through the Discovery Project
under Grant DP200100096. Simulations were undertaken with the assistance of resources and services from the National Computational Infrastructure (NCI), which is supported by the Australian Government.
}\\}
\IEEEauthorblockA{ECSE Department, Monash University, Clayton, VIC 3800, Australia\\
Email: \{tharaj.thaj, emanuele.viterbo\}@monash.edu}}

\maketitle
\begin{abstract}
This paper presents a linear complexity iterative rake detector for the recently proposed orthogonal time frequency space (OTFS) modulation scheme. The basic idea is to extract and coherently combine the received multipath components of the transmitted symbols in the delay-Doppler grid using maximal ratio combining (MRC) to improve the SNR of the combined signal. We reformulate the OTFS input-output relation in simple vector form by placing guard null symbols or zero padding (ZP) in the delay-Doppler grid and exploiting the resulting circulant property of the blocks of the channel matrix. Using this vector input-output relation we propose a low complexity iterative decision feedback equalizer (DFE) based on MRC. The performance and complexity of the proposed detector favorably compares with the state of the art message passing detector. An alternative time domain MRC based detector is also proposed for even faster detection. We further propose a Gauss-Seidel based over-relaxation parameter in the rake detector to improve the performance and the convergence speed of the iterative detection. We also show how the MRC detector can be combined with outer error-correcting codes to operate as a turbo DFE scheme to further improve the error performance. 
All results are compared with a baseline orthogonal frequency division multiplexing (OFDM) scheme employing a single tap minimum mean square error (MMSE) equalizer.
\end{abstract}

\begin{IEEEkeywords} 
  OTFS, Detector, Decoder, Rake, Maximal Ratio Combining,  Delay--Doppler Channel, Turbo, DFE, Gauss Seidel, Successive Over-Relaxation. 
\end{IEEEkeywords}
\section{Introduction}
Orthogonal time frequency and space (OTFS) is a new two dimensional (2D) modulation technique that transforms information symbols in the  delay-Doppler domain to the familiar time-frequency domain by spreading all the information symbols (e.g., QAM) over both time and frequency to achieve maximum {\em effective}\footnote{Effective diversity introduced for OTFS in \cite{effDiv} is a more meaningful measure of the actual diversity at practical SNR values, when the number of transmitted symbols is large.} diversity \cite{Hadani,effDiv}. As a result, a time-frequency selective channel due to multipath fading and mobility, is converted into a {\em separable} and {\em quasi-orthogonal} interaction, where all received information symbols experience roughly the same localized impairment \cite{Hadani}. Hence, for each information symbol, the received components in all the delay-Doppler diversity branches can be separated and coherently combined.

OTFS can also be interpreted as a two-dimensional code division multiple access (CDMA) scheme, where  information symbols are spread in both time and frequency, differently from conventional CDMA systems \cite{Hadani}. 
In direct sequence CDMA operating in a multipath fading channel, a rake receiver works by combining the delayed components (or echoes) of the transmitted symbols extracted by using matched filters tuned to the respective delay  shifts. Similarly, in the case of OTFS, the received delay shifted and Doppler shifted components of the transmitted information symbols can be extracted and coherently combined using linear diversity combining techniques to improve the SNR of the accumulated signal.

Diversity combining techniques are well studied in the literature starting from Brennan's paper on linear diversity combining \cite{MRC0}. Rake receivers for time domain combining using a variety of linear combining schemes like maximal ratio combining (MRC), equal gain combining (EGC) and selection combining (SC) are discussed in \cite{MRC01,MRC1}. MRC is shown to be optimal in the case of correlated and uncorrelated branches, even for unequal noise and interference power in the branches \cite{MRC2}. Moreover, iterative rake combining schemes and its variants are shown to combat inter-symbol interference better and are well investigated in the literature for CDMA systems \cite{GS1}.

In this paper, we propose an iterative rake receiver for the OTFS system using the maximal ratio combining scheme. Following \cite{WCNC_paper}, we group the delay-Doppler grid symbols into vectors according to their delay index and reformulate the input-output relation between the transmitted and received frames in terms of these transmitted and received vectors. By placing some null symbols (zero-padding (ZP)) in the delay-Doppler domain we arrive at a reduced input-output relation, which allows the use of the maximal ratio combining to design a low complexity detector for OTFS. The overhead of the null guard symbols, needed for the proposed detection scheme, also allows to insert pilot symbols at no additional cost \cite{Ravi3}. These null symbols in the delay-Doppler domain act as interleaved ZP guard bands in the time-domain. Taking advantage of this interleaved time-domain ZP, we further present an alternate low complexity time-domain MRC based detection for OTFS.

OTFS with the ZP guard band as mentioned above is similar to the Doppler-resilient orthogonal signal division multiplexing (D-OSDM) scheme recently proposed in \cite{D-OSDM} for under water acoustic channels \cite{UWA-channel} which is modelled as relatively faster time-varying as compared to the vehicular channel model assumption \cite{EVA}. Even though the information symbols in both schemes are transmitted in the delay-Doppler domain, the main advantage of the general OTFS transceiver structure is the provision to insert arbitrary frequency domain windowing, which is not a part of the D-OSDM scheme. Windowing allows OTFS to select a subset of sub-carriers for transmission and reception, which is particularly useful in multi-user communication schemes.

The rest of the paper is organized as follows. In Section II, we discuss the system model and derive the input-output relation in the vector form. To understand the operation of the proposed detector, we look at the input-output relation in delay-time and time domains in Section III. In Section IV, the proposed MRC based iterative rake detector, its low complexity implementation and the conditions for convergence are described. In Section V, we propose further improvements to the rake detector providing faster convergence and better error performance. The simulation results are given in Section VI followed by a discussion on the complexity of the proposed algorithm in Section VII. Section VIII contains our concluding remarks and future research directions.

\section{OTFS System Model}
\subsection{Notations}
The following notations will be followed in this paper: $a$, $\bf{a}$, ${\bf A}$ represent a scalar, vector, and matrix, respectively; ${\bf a}(n)$ and ${\bf A}(m,n)$ represent the $n$-th and $(m,n)$-th element of ${\bf a}$ and ${\bf A}$, respectively; ${\bf A}^\dag$, ${\bf A}^*$ and ${\bf A}^n$ represent the Hermitian transpose, complex conjugate and $n$-th power of ${\bf A}$. The set of $M \times N$ dimensional matrices with complex entries are denoted by ${\mathbb{C}}^{N \times M}$. Let $\circledast$ represent circular convolution, $\otimes$, the Kronecker product, $\circ$, the Hadamard product (i.e., the element wise multiplication) and, $\oslash$, the Hadamard division (i.e., the element wise division). Let $|\mathcal{S}|$ denote the cardinality of the set $\mathcal{S}$, $\mathrm{tr}(A)$, the trace of the square matrix ${\bf A}$, vec$({\bf A})$, the column-wise vectorization of the matrix ${\bf A}$ and ${\rm vec}_{N,M}^{-1}({\bf a})$ is the matrix formed by folding a vector ${\bf a}$ into a $N\times M$ matrix by filling it column wise. Let ${\bf F}_N$ be the normalized $N$ point discrete Fourier transform (DFT) matrix with elements ${\bf F}_N(i,k)=N^{-1/2}{\rm e}^{-j2\pi ik/N}$ and ${\bf F}_N^{\dag}$ the inverse discrete Fourier transform (IDFT) matrix, ${\bf I}_M$, the $M \times M$ identity matrix. The vectors ${\bf 0}_N$ and ${\bf 1}_N$ denote a $N$ length column vector of zeros and ones, respectively. The scalar $z={\rm e}^{\frac{j2\pi}{MN}}$. 
\subsection{Transmitter and Receiver Operation} 
The transmitter and receiver operations for the general OTFS system are described in \cite{Ravi2,farhang}. We will be using the following matrix/vector representation throughout the paper. Let ${\bf X}$, ${\bf Y}$ $\in \mathbb{C}^{M \times N}$ be the transmitted and received two-dimensional delay-Doppler grid, forming a {\em frame} of $M \times N$ $\mathcal{Q}$-QAM symbols, with unit average energy. 
Let ${\bf x}_m, {\bf y}_m \in \mathbb{C}^{N \times 1}$ be column vectors containing the symbols in the $m$-th row of ${\bf X}$ and ${\bf Y}$, respectively: 
${\bf x}_m$ = $[{\bf X}(m,0), {\bf X}(m,1), \cdots, {\bf X}(m,N-1)]^\text{T}$ and ${\bf y}_m$ = $[{\bf Y}(m,0), {\bf Y}(m,1), \cdots, {\bf Y}(m,N-1)]^\text{T}$, where $m$ and $n$ denote the delay (row) and Doppler (column) indices, respectively, in the two-dimensional grid. The total frame duration and bandwidth of the transmitted OTFS signal frame are $T_f=NT$ and $B = M \Delta f$, respectively. 
We consider the case where $T\Delta f=1$, i.e., the OTFS signal is critically sampled for any pulse shaping waveform.
 
\subsubsection{Basic OTFS Transmitter and Receiver}
The delay-Doppler domain symbols in ${\bf X}$ is converted to the time-frequency domain $({\bf X}_{\rm tf})$ using the inverse symplectic fast Fourier transform (ISFFT) operation.
\begin{equation}
    {\bf X}_{\rm tf}={{\bf F}} _M\cdot{{\bf X}}\cdot{{\bf F}} _N^{\dag}
    \label{Xtf}
\end{equation}
The “Heisenberg transform modulator” generates the time domain signal from the time-frequency samples using an M-point IFFT along with the pulse-shaping waveform $g_{\rm tx}(t)$. The transmitted signal can be written as
\begin{equation} {{\bf S}} = {{\bf G}} _{\rm tx}\cdot ({{\bf F}} _M^{\dag}\cdot{\bf X}_{\rm tf}) = {{\bf G}} _{\rm tx}\cdot({{\bf X}}\cdot{{\bf F}} _N^{\dag})\end{equation}
where the diagonal matrix ${\bf G}_{\rm tx}$ has the samples of $g_{\rm tx}(t)$ as its entries: ${\bf G}_{\rm tx}=\text{diag}[g_{\rm tx}(0),g_{\rm tx}(T/M),\ldots,g_{\rm tx}((M-1)T/M)]\in\mathcal{C}^{M\times M}$. Let $\thicktilde{\bf X}$
be the matrix containing the delay-time samples before applying pulse shaping waveform and is related to the delay-Doppler domain symbols as
\begin{align}&{\thicktilde{\bf X}}^{\rm T}=[\thicktilde{\bf x}_0, \ldots, \thicktilde{\bf x}_{M-1}]={{\bf F}} _N^{\dag}[{\bf x}_0, \ldots, {\bf x}_{M-1}]={{\bf F}} _N^{\dag}\cdot{\bf X}^{\rm T}.\label{OTFS_mod} \end{align} 

The time domain vector ${\bf s} \in \mathcal{C}^{NM \times 1}$, to be transmitted into the physical channel can be written as
 \begin{align}
    {\bf s}=\text{vec}({\bf G}_{\rm tx}\cdot\thicktilde{\bf X}).
    \label{dt_to_dd_tx}
\end{align}
These samples are pulse shaped and transmitted as a continuous time signal $s(t)$.
At the receiver, the delay-time samples are obtained from the  sampled received time domain waveform ${\bf r} \in \mathbb{C}^{NM\times 1}$ as
\begin{align}
    \thicktilde{\bf Y}=\text{vec}_{N,M}^{-1}\!\left(({\bf I}_M\otimes{\bf G}_{\rm rx})\cdot{\bf r}\right),
    \label{dt_to_dd_rx}
\end{align}
where the diagonal matrix ${\bf G}_{\rm rx}$ has the samples of $g_{\rm rx}(t)$ as its entries: ${\bf G}_{\rm rx}=\text{diag}[g_{\rm rx}(0),g_{\rm rx}(T/M),\ldots,g_{\rm rx}((M-1)T/M)]\in\mathcal{C}^{M\times M}$ is the pulse shaping filter at the receiver. The received delay-Doppler and delay-time domain symbols are related as
\begin{align}
&{{\bf Y}}^{\rm T}=[{\bf y}_0, \ldots, {\bf y}_{M-1}]={{\bf F}} _N[\thicktilde{\bf y}_0, \ldots, \thicktilde{\bf y}_{M-1}]={\bf F}_N\cdot\thicktilde{\bf Y}^{\rm T}. \label{dt2dd_rx} 
\end{align} 

\subsubsection{Rectangular pulse shaping waveforms}
\begin{figure*}
\centering
{\includegraphics[trim=17 0 0 10,clip,height=8.0in,width=7.2in]{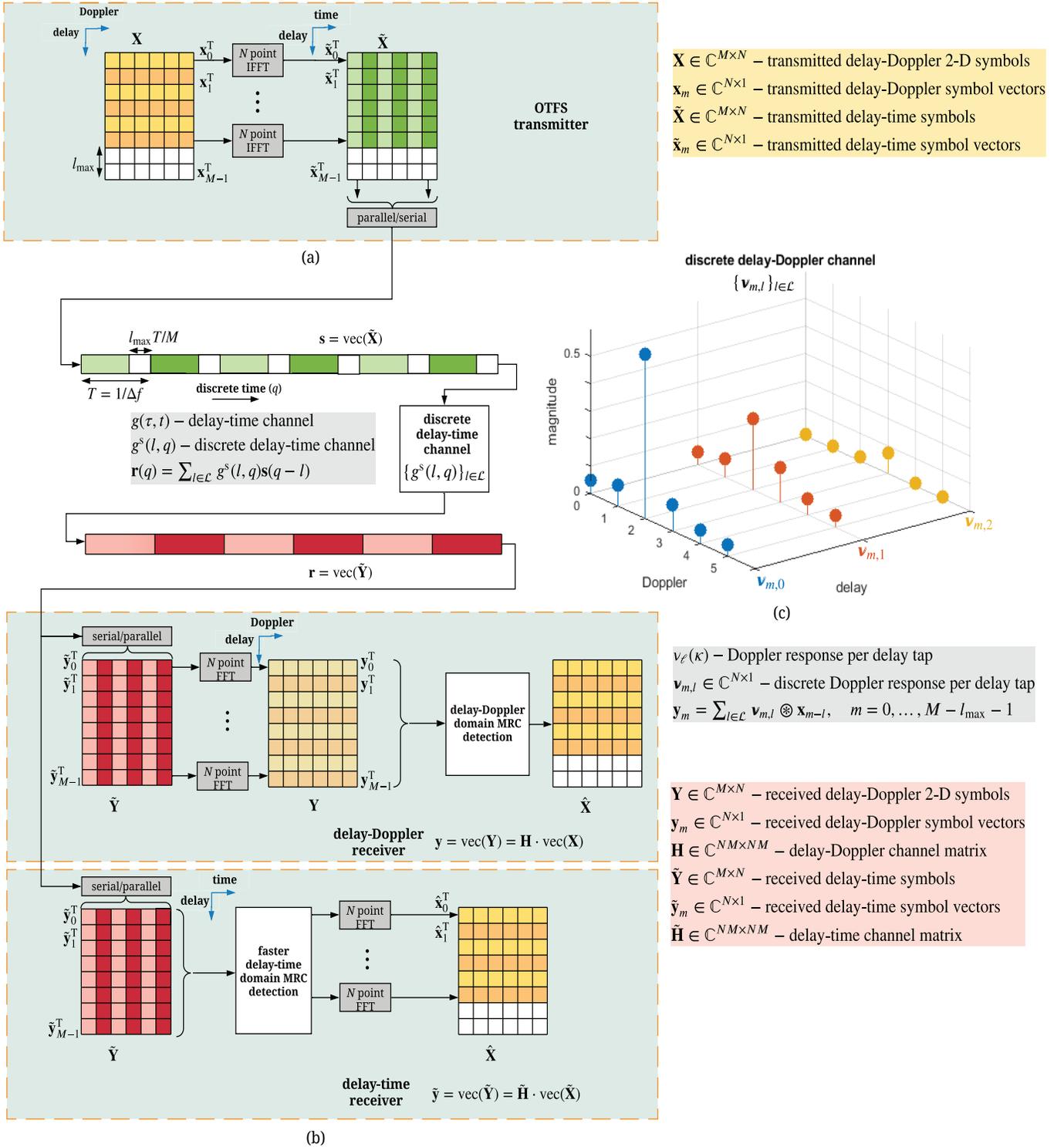}
 \vspace{-2mm}\caption{Discrete baseband model of the ZP-OTFS system for $N=6,M=8$ for (a) transmitter (b) receiver and (c) the discrete delay-Doppler channel at the set of discrete delay tap indices $\mathcal{L}=\{0,1,2\}.$ The samples shown using the same colour in (c) represent the Doppler response in the same delay tap. In (b), two versions of the proposed Rake receiver are presented (see Section IV). The receiver chain on the top part of (b) operates directly in the information symbol domain, i.e., the delay-Doppler domain (see Algorithm 1 in Section IV.A) and the bottom part of (b) is the faster version (see Algorithm 2 in Section IV.B) which operates in the delay-time domain.} 
\label{OTFSsys}}
\end{figure*}
In this paper, we consider  rectangular transmit and received pulse shaping waveforms which is equivalent to time-domain windowing, i.e.,  ${\bf G}_{\rm tx}={\bf G}_{\rm rx}={\bf I}_M$.\footnote{In general, the pulse shaping waveforms $({\bf G}_{\rm tx})$ could be circulant matrices (equivalent to time-domain filtering).}
The transmitted and received time domain discrete samples ${\bf s,r}$ can then be written in terms of the delay-time samples $\thicktilde{\bf x}_m $ and $\thicktilde{\bf y}_m$ as 
\begin{align}
    {\bf s}(m+nM)=\thicktilde{\bf x}_m(n), \nonumber \\
    {\bf r}(m+nM)=\thicktilde{\bf y}_m(n).
    \label{r2y_relation}
\end{align}
 In this case, the transmitted and received discrete time domain signal samples can be related to the delay-Doppler domain information symbols as
\begin{align}
    &{\bf s}=\text{vec}({\bf X}\cdot{\bf F}_N^{\dag}) \quad\text{and}\quad{\bf r}=\text{vec}({\bf Y}\cdot{\bf F}_N^{\dag}). \label{Zak}
\end{align}
The operation in (\ref{Zak}) in the literature is known as the inverse discrete Zak transform \cite{Hadani2}.

 The simplified transmitter and receiver baseband equivalent model for rectangular pulse shaping waveforms and {\em two} MRC based detection methods (to be discussed in Section IV) are shown in Fig. \ref{OTFSsys} (a) and (b). The last $l_{\rm max}$ symbol vectors (rows) of the transmitted delay-Doppler grid, where $l_{\rm max}$ is the maximum channel delay spread index, are made zero to avoid inter-block interference in the time-domain. These zero vectors aid in reducing the complexity of detection for OTFS (explained in Section III-B) by allowing parallel processing of the $N$ independent time domain blocks of duration $T$. 
 
 For the rest of the paper, to differentiate with the basic OTFS scheme, as discussed in \cite{Hadani,Ravi2}, we refer to the above scheme including zero padding as the ZP-OTFS. Our main motivation behind adding the delay-Doppler domain ZP is the design of a low complexity detector for OTFS, \cite{WCNC_paper}. Adding a ZP along the delay dimension in the OTFS delay-Doppler grid can be seen as analogous to the time-domain CP or ZP added in orthogonal frequency division multiplexing (OFDM), which allows the design of a single tap equalizer in the time-frequency domain, and hence contribute to reduction in detector complexity. Moreover, in OTFS, the ZP can be used as guard band for the pilot in the delay-Doppler domain \cite{Ravi3}, and hence reduction in detector complexity can be achieved at little cost, which is convenient for the ZP-OTFS system.

\subsection{Continuous Time Baseband Channel Model}

Consider a baseband equivalent channel model\footnote{We do not consider the effects of carrier frequency and antenna gains in this paper.} with $P$ propagation paths, where $h_i$ is the complex {\em path gain}, $\ell_i$ and $\kappa_i$ are the {\em normalized delay shift} and {\em normalized Doppler shift}, respectively, associated with the $i$-th path, where $\ell_i,\kappa_i \in \mathbb{R}$ are not necessarily integers. 
The actual delay and Doppler shift for the $i$-th path is given by $\tau_i=\frac{\ell_i}{M\Delta f}< \tau_{\max}=\frac{\ell_{\max}}{M\Delta f}$, $\nu_i=\frac{\kappa_i}{NT}$ with $|\nu_i|<\nu_{\max}$. 
We assume that the channel is {\em under-spread}, i.e., $\tau_{\max}{\nu_{\max}} \ll 1$. Under the under-spread assumption,  $\ell_{\max}<M$ and the normalized Doppler shifts $-N/2<\kappa_i<N/2$. Since the number of channel coefficients $P$ in the delay-Doppler domain is typically limited, the channel response has a sparse representation \cite{Hadani,Ravi2}:
\begin{equation} 
\label{eq:channel}
h(\tau, \nu) = \sum _{i=1}^{P} h_i \delta (\tau -\tau _i) \delta (\nu -\nu _i).  
\end{equation}
Alternatively, we can write,
\begin{equation} 
\label{eq:channel1}
h(\tau, \nu
) = \sum _{\ell \in \mathcal{L}^{\prime}} \sum _{\kappa \in \mathcal{K}_{\ell}}  {\nu}_{\ell}(\kappa) \delta (\tau -{\ell}T/M) \delta (\nu -{\kappa}{\Delta f}/N)  
\end{equation}  
where  $\mathcal{L}^{\prime}=\{\ell_i \}$ is the set of $L^{\prime}=|\mathcal{L}^{\prime}|$ distinct {\em normalized delay shifts} among the  $P$ paths in the delay-Doppler domain, $\mathcal{K}_{\ell}=\{ \kappa_i\mid \ell=\ell_i\}$ is the set of {\em normalized Doppler shifts} for each path with {\em normalized delay shift} $\ell_i$, and 
 \begin{align} {\nu}_{\ell}(\kappa) = \left. \left\lbrace \begin{array}{ll}  h_i, & \text{if } \ell=\ell_i \text{ and } \kappa=\kappa_i   \\ 0, & \text{otherwise.} \end{array}\right. \right.
 \label{4} 
 \end{align}
is the $\ell$-th delay tap {\em  Doppler response}. The magnitude of a {\em  Doppler response} function ${\nu}_{\ell}(\kappa)$ evaluated at integer delay and Doppler shifts is shown in Fig. \ref{OTFSsys}.

The corresponding continuous time-varying channel impulse response function can be written, for all $\ell \in \mathcal{L}^{\prime}$, as
\begin{align} 
g(\tau,t)&= \int_\nu h(\tau,\nu){\rm e}^{j2\pi\nu (t-\tau)}\, d\nu \label{gttau}.
\end{align}
Substituting (\ref{4}) into (\ref{gttau}) and evaluating (\ref{gttau}) at $\tau=\ell T/M$, we get,
\begin{align}
g(\ell T/M,t)&=\sum_{\kappa \in \mathcal{K}_{\ell}} {\nu}_{\ell}(\kappa){\rm e}^{j2\pi\kappa\frac{\Delta f}{N} (t-\ell T/M)} 
\label{td_cont}
\end{align} 
which represents the {\em delay-time channel response}, for all $\ell \in \mathcal{L}^{\prime}$.

\subsection{Discrete Time  Baseband Channel Model}
At the transmitter, the OTFS frame of bandwidth $B=M\Delta f$ is up-converted to a carrier frequency $f_c$ to occupy a pass band channel, assuming $f_c \gg B$.  At the receiver, the channel impaired signal is down-converted to baseband and sampled at $M\Delta f$ Hz, thereby limiting the received waveform to $NM$ complex samples. Therefore, from a communication system design point of view, it is convenient to have a discrete baseband equivalent representation of the system, \cite{Wireless_book}. 

In the previous section, we looked at the continuous time model of the channel. The discrete time model is obtained by sampling the received waveform $r(t)$ at sampling intervals $t=qT/M$, where $0 \leq q \leq NM-1$, which discretizes the delay-time channel. The set of {\em normalized delay shifts}, $\mathcal{L}^{\prime}$ is therefore replaced as $\mathcal{L}$ with the set of $L=|\mathcal{L}|$ discrete delay taps representing delay shifts at integer multiples of the sampling period $T/M$. Recall that $\frac{\Delta f}{N}$ and $\frac{T}{M}$ are the Doppler and delay resolution, respectively, of the delay-Doppler grid, given $T{\Delta f}=1$. Following from the sampling theorem \cite{Wireless_book}, the discrete baseband delay-time channel model of (\ref{td_cont}) is given as, 
\begin{align}
g^{\rm s}(l,q)&=\sum_{\ell \in \mathcal{L}^{\prime}} \left(\sum_{\kappa \in \mathcal{K}_{\ell}}{\nu}_{\ell}(\kappa)z^{\kappa(q-l)}\right) {\rm sinc}(l-\ell)
\label{t_eq11}\end{align}
where ${\rm sinc}(x)={\rm sin}(\pi x)/{(\pi x)}$ and $z={\rm e}^{\frac{j2\pi}{NM}}$.

Note that, due to fractional delays, the sampling at the receiver introduces interference between Doppler responses at different delay shifts. This is due to sinc reconstruction of the delay-time response at fractional delay points ($\ell$), \cite{Wireless_book}. However, under the assumption that the channel delay shifts can be modelled as integer delay shifts without loss of accuracy, i.e., when $\mathcal{L}^{\prime}=\mathcal{L}$ and hence $\ell=l' \in \mathbb{Z}$, the sinc function in (\ref{t_eq11}) reduces to
\begin{align} {\rm sinc}(l-l') = \left. \left\lbrace \begin{array}{ll}  1, & \text{if } l'=l  \\ 0, & \text{otherwise.} \end{array}\right. \right.
 \end{align}
Consequently, the relation between the actual Doppler response and the sampled time-domain channel at each integer delay tap $l \in \mathcal{L}$ in (\ref{t_eq11}) reduces to
\begin{equation}
g^{\rm s}(l,q)=\sum_{\kappa \in \mathcal{K}_l} {\nu}_{l}(\kappa)z^{\kappa(q-l)}.
\label{t_eq}\end{equation}
Here we want to remind the readers that the effective channel as seen by the receiver depends on the actual channel response as well as the operation parameters (delay and Doppler resolution) of the receiver. 

For the rest of the paper, to clearly differentiate between the real continuous channel and the effective discrete channel as seen by the receiver, we use $\ell$ and $\kappa$ to denote the normalized delay and Doppler shifts (not necessarily integers) associated with the channel whereas $l$ and $k$ is used only to denote integer delay and Doppler shift indices, respectively, associated with the channel sampled on the OTFS delay-Doppler grid. 
\subsection{Input-Output Relations in Delay-Doppler Domain \label{Sec:IOrelationDD}}

 In this section, we reformulate the input-output relation with rectangular pulse shaping waveforms, for the ZP-OTFS system shown in Fig. \ref{OTFSsys}.

Starting from the received time-domain signal $r(t)$, the continuous time domain input-output relation can be written as 
\begin{equation}
    r(t)=\int_{0}^{\tau_{\rm max}}g(\tau,t)s(t-\tau)\, d\,\tau.
\end{equation}

From (\ref{t_eq11}), the corresponding discrete time-domain input-output relation when the transmitted and received time-domain signals are sampled at $t=qT/M$ can be written as

\begin{equation}
    {\bf r}(q)=\sum_{l \in \mathcal{L}}g^{\rm s}(l,q){\bf s}(q-l) \label{disc_time}
\end{equation}
where ${\bf r}(q)=r(q\frac{T}{M})$, ${\bf s}(q)=s(q\frac{T}{M})$. Using the relations in (\ref{r2y_relation}), we split the time index $q=0, \ldots, MN-1$ in terms of the delay and Doppler frame indices as $q=(m+nM)$, where the $m=0, 1, \ldots, M-1$ and $n=0, 1, \ldots, N-1$. Then replacing $\thicktilde{\pmb \nu}_{m,l}(n)=g^{\rm s}(l,m+nM)$, we can rewrite (\ref{disc_time}) in terms of the delay-time symbol vectors as
\begin{align}
    {\thicktilde{\bf y}}_m(n)&=\sum_{l \in \mathcal{L}}{\thicktilde{\pmb \nu}_{m,l}(n)}{\thicktilde{\bf x}}_{m-l}(n)\label{disc_time3}
\end{align}
 where  $\thicktilde{\pmb \nu}_{m,l} \in \mathbb{C}^{N \times 1}$ is given as
\begin{align}
   {\thicktilde{\pmb \nu}_{m,l}(n)}&=
   \sum_{\ell \in \mathcal{L}^{\prime}}\left(\sum_{\kappa \in \mathcal{K}_l} {\nu}_{\ell}(\kappa) z^{\kappa(m-l)}{\rm e}^{\frac{j2\pi\kappa n}{N}}\right){\rm sinc}(l-\ell)
   \label{g_lt_to_v_ml2}.
\end{align}
 For integer delay tap channel assumption, i.e., $l=\ell \in \mathbb{Z}$, (\ref{g_lt_to_v_ml2}) becomes, 
\begin{align}
   {\thicktilde{\pmb \nu}_{m,l}(n)}&=
   \sum_{\kappa \in \mathcal{K}_l} {\nu}_{l}(\kappa) z^{\kappa(m-l)}{\rm e}^{\frac{j2\pi\kappa n}{N}}
   \label{g_lt_to_v_ml22}.
\end{align}
We can note from (\ref{g_lt_to_v_ml22}) that the discrete delay-time response ${\thicktilde{\pmb \nu}_{m,l}(n)}$ for each delay tap $l$ at time instants $t=\frac{m}{M}T+nT$ is related to the inverse Fourier transform of the {\em Doppler response} ${\nu}^{\rm}_{l}(\kappa)$ of the $l$-th delay tap sampled at time $t=\frac{m}{M}T$.  
We may ignore the case in (\ref{disc_time3}) when $m-l<0$ i.e., when there is inter-block interference due to channel delay spread, by making $\thicktilde{\bf x}_m(n)=0$ for all $n$ when $m-l<0$ such that, 
    \begin{align}
    {\thicktilde{\pmb \nu}_{m,l}}(n) \thicktilde{{\bf x}}_{m-l}([n-k]_{N})=0, \text{ if } m < l
    \label{condition1}
    \end{align} 
This is equivalent to placing null symbol vectors ${\bf 0}_N$ in the last $l_{\max}$ rows of ${\bf X}$ (zero padding along the delay dimension of the OTFS grid). Hence, we can set, for $n=0, \ldots, N-1$,
\begin{align}
{\bf x}_m(n)={\thicktilde{\bf x}}_m(n)=0, \text{ if } m \geq M-l_{\max} \label{cond2}
\end{align}

The delay-Doppler domain received symbols can be obtained by taking an $N$-point FFT of the delay-time received symbol vectors (\ref{dt2dd_rx})
\begin{align}
    {{\bf y}}_m={\bf F}_N\cdot\thicktilde{\bf y}_m=&\sum_{l \in \mathcal{L}}{\bf F}_N\cdot({\thicktilde{\pmb \nu}_{m,l}}\circ{\thicktilde{\bf x}}_{m-l}) \nonumber \\
    &= \sum_{l \in \mathcal{L}}({\bf F}_N\cdot{\thicktilde{\pmb \nu}_{m,l}})\circledast({\bf F}_N\cdot{\thicktilde{\bf x}}_{m-l})
    \nonumber \\ &= \sum_{l \in \mathcal{L}}{\pmb \nu}_{m,l}\circledast{\bf x}_{m-l}
    \label{9}
\end{align}
where,
\begin{align}
    {{\pmb \nu}_{m,l}(k)}=\frac{1}{\sqrt{N}}\sum_{n=0}^{N-1}\thicktilde{\pmb \nu}_{m,l}(n){\rm e}^{\frac{-j2\pi kn}{N}}
    \label{g_lt_to_v_ml3}
\end{align}
for $0\leq k\leq N-1$, $0\leq m < M-l_{max}$, is the discrete {\em Doppler spread vector} in the $l$-th channel delay tap, experienced by all the symbols in the $\left(m-l\right)$-th row of the $M\times N$ OTFS delay-Doppler grid. Fig. \ref{OTFSsys} (c) shows the discrete Doppler spread vectors ${\pmb \nu}_{l,l}$ for ${\bf x}_0$. Substituting (\ref{t_eq11}), (\ref{g_lt_to_v_ml22}) and (\ref{g_lt_to_v_ml2}) in (\ref{g_lt_to_v_ml3}), we can write the discrete Doppler spread vector ${\pmb \nu}_{m,l} \in \mathbb{C}^{N \times 1}$ in terms of the channel {\em Doppler response} ${\nu}_{\ell}(\kappa)$, for a channel model assuming:
\subsubsection{Fractional delay and fractional Doppler shifts}
\begin{align}
{{\pmb \nu}_{m,l}(k)}&=\frac{1}{\sqrt{N}}\sum_{\ell \in \mathcal{L}^{\prime}}\left(\sum_{\kappa\in \mathcal{K}_\ell}{\nu}_{\ell}(\kappa)z^{\kappa(m-l)}\zeta_N(\kappa-k)\right) {\rm sinc}(l-\ell)\label{g_lt_to_v_ml4}\end{align}
where $\ell, \kappa \in \mathbb{R}$ and the {\em periodic sinc function} $\zeta(\cdot)$ includes the extra phase and magnitude variations in the Doppler spread vectors due to fractional Doppler shifts, given as
\begin{align}
     \zeta_N(x)=\frac{1}{\sqrt{N}}\sum_{n=0}^{N-1}{\rm e}^{\frac{j2\pi x n}{N}}=\frac{1}{\sqrt{N}}\frac{{\rm sin}\left(\pi x\right)}{{\rm sin}(\pi x/N)}{\rm e}^{\frac{j\pi x(N-1)}{N}}
\end{align}
\subsubsection{Integer delay and fractional Doppler shifts}
For integer values of $(l-\ell)$, the function  ${\rm sinc}(l-\ell)$ evaluates to $1$ when $l=\ell$ and {\em zero} else where. Hence (\ref{g_lt_to_v_ml4}) reduces to
\begin{align}
{{\pmb \nu}_{m,l}(k)}&=\frac{1}{\sqrt{N}}\sum_{\kappa\in \mathcal{K}_l} {\nu}_{\ell}(\kappa)z^{\kappa(m-l)}\zeta_N(\kappa-k)\label{g_lt_to_v_ml5}\end{align}
for $l=\ell \in \mathbb{Z}$ and $\kappa \in \mathbb{R}$
\subsubsection{Integer delay and integer Doppler shifts }
For integer values of $x$, the function  $\zeta_N(x)$ evaluates to $\sqrt{N}$ when $x=0$ and {\em zero} else where. Hence (\ref{g_lt_to_v_ml5}) reduces to the simple form
 \begin{align} {{\pmb \nu}_{m,l}(k)} = \left. \left\lbrace \begin{array}{ll}  {\nu}_{\ell}(\kappa)z^{\kappa(m-l)}, & \text{if } l=\ell \text{ and } k=[\kappa]_N  \\ 0, & \text{otherwise.} \end{array}\right. \right.
\label{g_lt_to_v_ml6} \end{align}
for $\ell, \kappa \in \mathbb{Z}$.

{\em Remark} -- The above {\rm three} cases result in phase changes $z^{\kappa(m-l)}$ due to the rectangular pulse shaping waveforms. For the ideal pulse shaping waveform assumption, it was shown in \cite{WCNC_paper,Ravi2} that the Doppler spread vectors ${\pmb \nu}_{m,l}$ are invariant on the 2-D delay-Doppler grid and hence not dependent on the row index $m$. The phase variations $z^{\kappa(m-l)}$ can be ignored in (\ref{g_lt_to_v_ml4}), (\ref{g_lt_to_v_ml5}) and (\ref{g_lt_to_v_ml6}). As a result (\ref{9}) is a simple time-invariant 2-D circular convolution as shown in \cite{WCNC_paper,Ravi2}. 
It is important to note that ignoring such phase variations in the detection process results in significant performance degradation.
$\square$ 

For the rest of the paper and simulations, we assume integer delays and fractional Doppler shifts for rectangular pulse shaping waveforms, i.e., we consider the discrete input-output relation of the form given in (\ref{9}) and (\ref{g_lt_to_v_ml5}) where $\mathcal{L}^{\prime}=\mathcal{L} \in \mathbb{Z}$.

\begin{figure}
\centering
{\includegraphics[trim=3 5 0 5,clip,height=2.2in,width=3.4in]{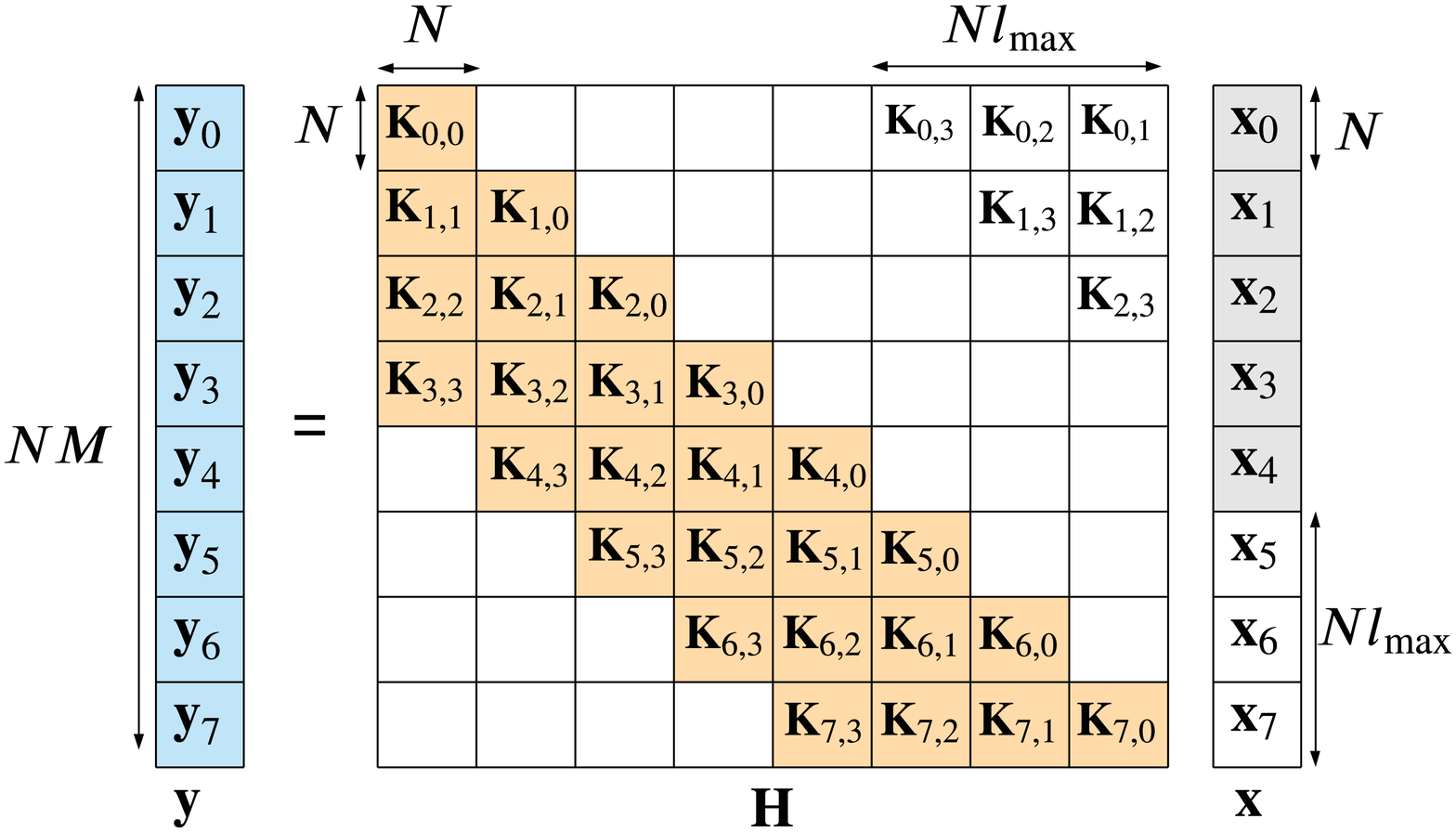}
 \vspace{-2mm}\caption{The delay-Doppler domain input-output relation ${\bf y}={\bf H}\cdot{\bf x}$ after adding null symbols only contains the shaded blocks for $N=M=8$ and $l_{max}=3$.}
\label{mat2}}
\end{figure}

 The OTFS delay-Doppler domain discrete system for the ZP OTFS system can be expressed in the matrix form as
\begin{equation}
    {\bf y}={\bf H}\cdot{\bf x}+{\bf w};
     \label{H_matrix}
\end{equation}

where  ${\bf x,y,w} \in { \mathbb{C}}^{NM \times 1}$ and ${\bf H} \in { \mathbb{C}}^{NM \times NM}$ is the OTFS channel matrix when transmitted and received symbol-vectors, ${\bf x}_m,{\bf y}_m \in { \mathbb{C}}^{N \times 1}$ are grouped and stacked as  ${\bf y}=[{\bf y}_0^\text{T}, {\bf y}_1^\text{T}, \cdots, {\bf y}_{M-1}^\text{T}]^\text{T}$, ${\bf x}=[{\bf x}_0^\text{T}, {\bf x}_1^\text{T}, \cdots, {\bf x}_{M-1}^\text{T}]^\text{T}$ and ${\bf w}=[{\bf w}_0^\text{T}, {\bf w}_1^\text{T}, \cdots, {\bf w}_{M-1}^\text{T}]^\text{T}$ is independent and identically distributed (iid) additive white guassian noise (AWGN) with variance $\sigma_w^2$. Referring to the vectorized form shown in Fig. \ref{mat2}, we convert the circular convolution between two vectors into the product of a  matrix and a vector by defining ${\bf K}_{m,l} \in \mathbb{C}^{N \times N}$ to be a banded  matrix for $ l \in \mathcal{L}$ and an all zero matrix otherwise
\begin{align*} {\bf K}_{m,l} & = 
  \text{circ}[{\pmb{\nu}}_{m,l}(0),\cdots,{\pmb{\nu}}_{m,l}(N-1)]\\[1ex]
  & = \left[\begin{array}{cccc} {{\pmb{\nu}}_{m,l}}(0) & {{\pmb{\nu}}_{m,l}}(N-1) & \cdots & {{\pmb{\nu}}_{m,l}}(1)\\ {{\pmb{\nu}}_{m,l}}(1) & {{\pmb{\nu}}_{m,l}}(0) & \cdots & {{\pmb{\nu}}_{m,l}}(2)\\ \vdots & \ddots & \ddots & \vdots \\ {{\pmb{\nu}}_{m,l}}(N-1) & {{\pmb{\nu}}_{m,l}}(N-2) & \cdots & {{\pmb{\nu}}_{m,l}}(0) \end{array}\right] ~.
\end{align*}
We note that the band width of each submatrix ${\bf K}_{m,l}$ of ${\bf H}$ is equal to the maximum Doppler spread $k_{\max}\leq N/2$  and the full channel matrix ${\bf H}$ has a band width equal to $N(l_{\max}+1)$. We can then write (\ref{9}) as 
     \begin{align} {{\bf y}}_m = \sum _{l \in \mathcal{L}} {\bf K}_{m,l}\cdot{{\bf x}}_{m-l}. \label{11} 
     \end{align}
Note that ${\bf K}_{m,l}$ (or ${\pmb{\nu}}_{m,{l}}$) can be considered as the linear time-variant channel between the receiver grid delay index $m$ and transmitter grid delay index $m-l$ in the OTFS delay-Doppler grid. Now (\ref{9}) and (\ref{11}) gives us a very simple equation relating the transmitted and received symbol-vectors that we defined at the start of this section. 

\section{Input-Output Relation in Other Domains}
 In this section, we discuss the ZP-OTFS input-output relation between the transmitted and received delay-time symbol vectors and discuss the advantages of carrying out significant part of the OTFS receiver processing in the delay-time domain. We also highlight some properties of the delay-time and time-domain channel matrices to later analyze the convergence of the proposed detector. 

When $N$ and $M$ are sufficiently large, considering the channel normalized delay and Doppler shifts ($\kappa_i$ and $\ell_i$) as integers has negligible effect on the accuracy of the channel representation. However, the effect of fractional Doppler is more pronounced for short OTFS frames, \cite{Ravi}. When $N$ is small, a single path with fractional Doppler shift is seen as a cluster of paths with integer Doppler shifts at the receiver. Depending on the resolution, more channel coefficients along the Doppler dimension are required to fully represent the channel state information needed for accurate detection at the receiver, \cite{Ravi2}. This increases the total number of paths $P$ for the discrete channel. To mitigate such problem, the value of $N$ may be increased, which, in turn, will increase the frame duration $NT$. However, the frame duration is limited by the {\em delay-Doppler coherence time},\footnote{This coherence time should not be confused with the traditional notion  related to the inverse of the Doppler spread,  \cite{WCNC_paper}.} i.e, the time over which the delay-Doppler channel coefficients remain constant. 

Another way of solving the fractional Doppler issue is by  dealing with the delay-Doppler channel coefficients in the delay-time domain. As Doppler shifts cannot be resolved in this domain, the number of delay-time channel coefficients is neither affected by the fractional Doppler shifts nor by the Doppler spread of that delay tap. Therefore, to fully take advantage of the OTFS performance in a rich Doppler spread regime (i.e., large $|\mathcal{K}_l|$'s), it is convenient to design a receiver with low complexity that is independent of the Doppler spread. 
\subsection{Delay-Time Domain}
  For the purpose of delay-time detection analysis in Section IV, we look at the matrix representation of the delay-time input-output relation. The  matrices ${\bf K}_{m,l}$ in the delay-Doppler domain can be diagonalized to $\thicktilde{\bf K}_{m,l}$ in the corresponding Fourier domain (delay-time domain) as
\begin{align*} 
&{\bf K}_{m,l} = {\bf F}_N\cdot\thicktilde{\bf K}_{m,l}\cdot{\bf F}_N^{\dag},\\
\implies&\thicktilde{\bf K}_{m,l}  = 
  \text{diag}[\thicktilde{\pmb{\nu}}_{m,{l}}(0),\cdots,\thicktilde{\pmb{\nu}}_{m,{l}}(N-1)]\\[1ex]
 &\text{where } \thicktilde{\pmb \nu}_{m,l}={\bf F}_N^{\dag}{\pmb \nu}_{m,l}
\end{align*}
\begin{figure}
\centering
{\includegraphics[trim=3 10 0 5,clip,height=2.2in,width=3.6in]{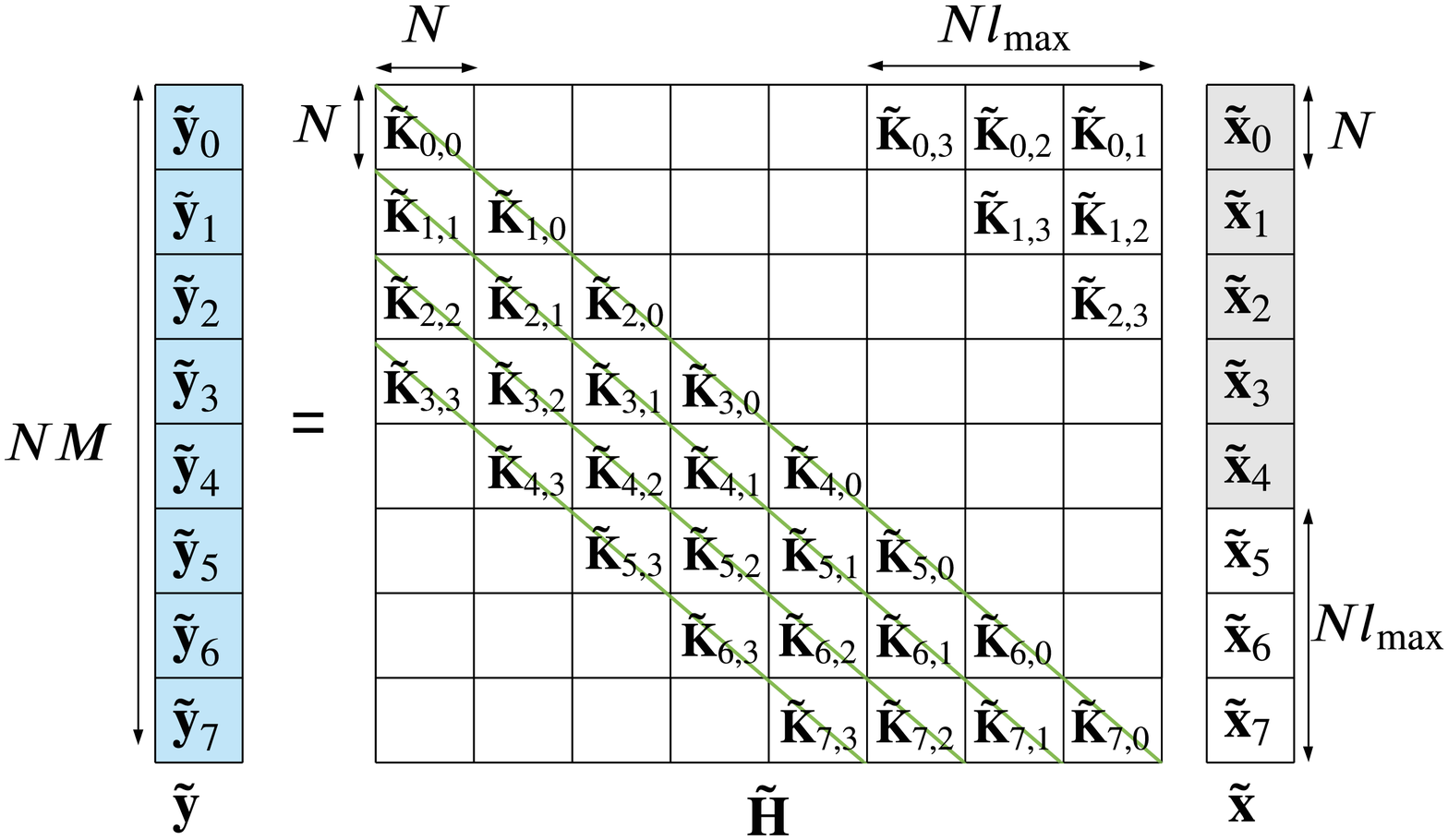}
 \vspace{-6mm}\caption{The delay-time domain input-output relation ($\thicktilde{\bf y}=\thicktilde{\bf H}\cdot\thicktilde{\bf x}$) after adding null symbols for $N=M=8$ and $l_{\max}=3$.}
\label{mat3}}
\end{figure}
thereby transforming the delay-Doppler domain channel matrix ${\bf H}$  into the delay-time domain channel matrix $\thicktilde{\bf H}$  by replacing the sub-matrices ${\bf K}_{m,l}$ in ${\bf H}$ with $\thicktilde{\bf K}_{m,l}$. Given the input-output relation in (\ref{H_matrix}) was simplified in (\ref{11}) by placing null symbols in the delay-Doppler grid as given in (\ref{cond2}), the strictly upper triangular blocks of $\thicktilde{\bf H}$ can also be set to zero.
The input-output relation in the delay-time domain, illustrated  in Fig. \ref{mat3}, can then be written in the matrix form as 
\begin{equation}
    \thicktilde{\bf y}=\thicktilde{\bf H}\cdot\thicktilde{\bf x}+\thicktilde{\bf w};
     \label{time_delay_io}
\end{equation}
where 
\begin{align}
&\thicktilde{\bf y}=({\bf I}_M\otimes{\bf F}_N^{\dag})\cdot{\bf y},\quad\thicktilde{\bf x}=({\bf I}_M\otimes{\bf F}_N^{\dag})\cdot{\bf x},\nonumber\\
&\thicktilde{\bf H}=({\bf I}_M\otimes{\bf F}_N^{\dag})\cdot{\bf H}\cdot({\bf I}_M\otimes{\bf F}_N),
\label{Htd}\end{align}
and $\thicktilde{\bf w}$ is the time domain AWGN vector. In this domain, the complexity of matrix multiplication is significantly reduced as the sparsity $L/N$ of $\thicktilde{\bf H}$  is less than or equal to the sparsity  $P/N$ of ${\bf H}$, where $L$ is the number of unique delay taps and $P$ is the total number of propagation paths. The delay-time domain channel matrix $\thicktilde{\bf H}$ is a banded block matrix (with a bandwidth of $Nl_{\max}+1$), where   $\thicktilde{\bf K}_{m,l} \in \mathcal{C}^{N \times N}$ are non-zero diagonal matrices for $m \geq l$ and $l \in \mathcal{L}$ and zero matrices otherwise. Consequently,  the delay-Doppler domain input-output relation in (\ref{9}) becomes
     \begin{align} {\thicktilde{\bf y}}_m = \sum _{l \in \mathcal{L}} \thicktilde{\pmb \nu}_{m,l}\circ{\thicktilde{\bf x}}_{m-l},\quad  \thicktilde{\bf x}_m={\bf 0}_N \text{ for } m \geq M-l_{\rm max}. \label{td_ioeq} 
     \end{align}
in the delay-time domain, where  $\thicktilde{\bf x}=[\thicktilde{\bf x}_0^\text{T}, \cdots, \thicktilde{\bf x}_{M-1}^\text{T}]^\text{T}$ and $\thicktilde{\bf y}=[\thicktilde{\bf y}_0^\text{T}, \cdots, \thicktilde{\bf y}_{M-1}^\text{T}]^\text{T}$.
    
\subsection{Time Domain}
\begin{figure}
\centering
{\includegraphics[trim=3 8 0 5,clip,height=2.2in,width=3.6in]{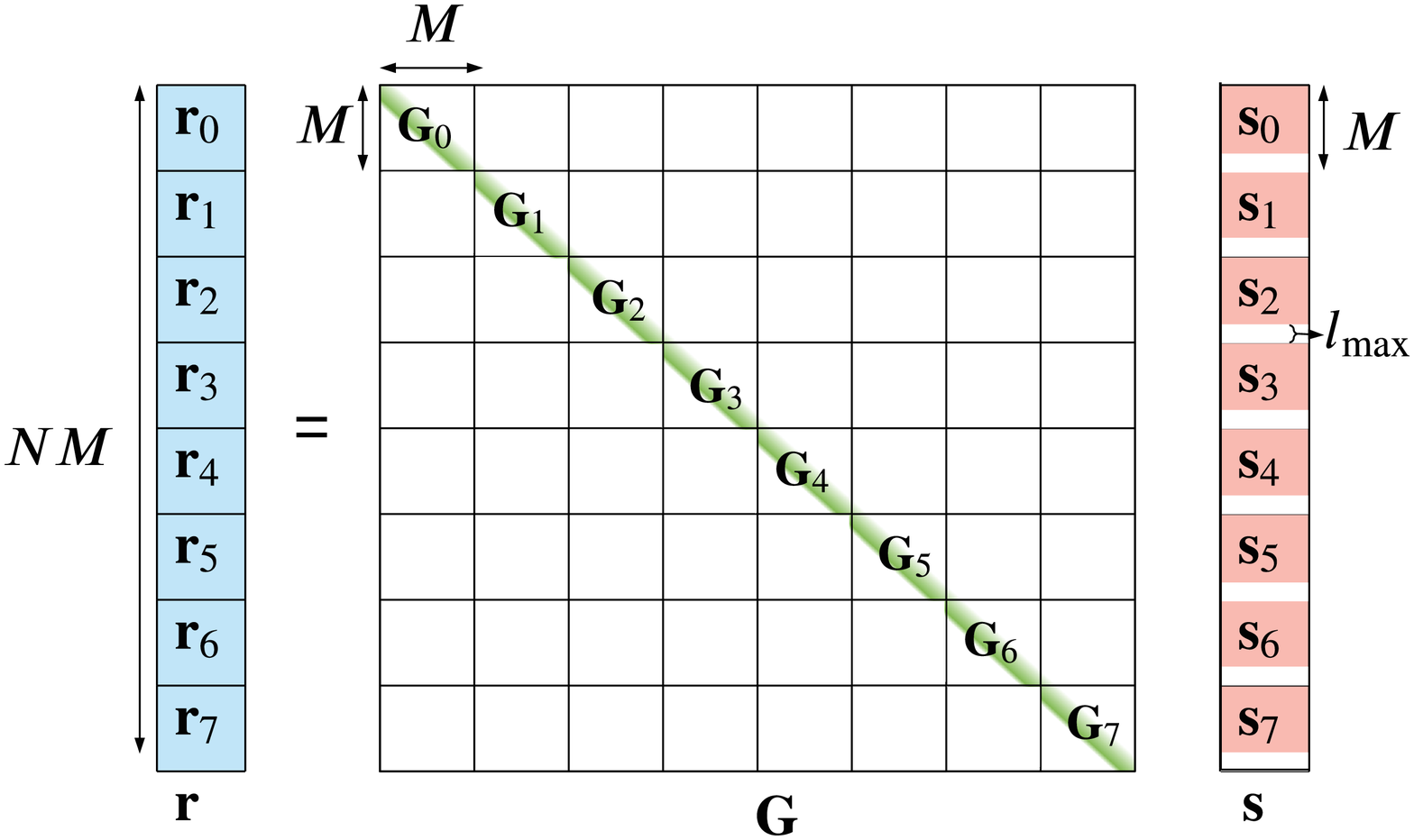}
 \vspace{-6mm}\caption{The time-domain input-output relation ${\bf r}={\bf G}\cdot{\bf s}$ after shuffling the matrix $\thicktilde{\bf H}$ as ${\bf G}={\bf P}\cdot\thicktilde{\bf H}\cdot{\bf P}^{\rm T}$ for $N=M=8$ and $l_{\max}=3$.}
\label{mat4}}
\end{figure}
 Here, we show how the time domain input-output relation is connected to the delay-Doppler and the delay-time domain input-output relations. 

From (\ref{r2y_relation}), it can be seen that the delay-time vectors $\thicktilde{\bf x}$ and $\thicktilde{\bf y}$ in (\ref{time_delay_io}) are simply shuffled versions of the time domain transmitted and received vectors ${\bf s}$ and ${\bf r}$, respectively. 
Let ${\bf s}$ and ${\bf r}$ be split into $N$ blocks each of size $M$, such that ${\bf s}=[{\bf s}_0^\text{T}, \cdots, {\bf s}_{N-1}^\text{T}]^\text{T}$ and ${\bf r}=[{\bf r}_0^\text{T}, \cdots, {\bf r}_{N-1}^\text{T}]^\text{T}$. 
Then $\thicktilde{\bf x}_m=[{\bf s}_0(m), \cdots, {\bf s}_{N-1}(m)]^\text{T}$ and $\thicktilde{\bf y}_m=[{\bf r}_0(m), \cdots, {\bf r}_{N-1}(m)]^\text{T}$. 

Let 
\begin{align} {\bf P}
  & = \left[\begin{array}{cccc} {\bf E}_{1,1} & {\bf E}_{2,1} & \cdots & {\bf E}_{M,1}\\ {\bf E}_{1,2} & {\bf E}_{2,2} & \cdots & {\bf E}_{M,2}\\ \vdots & \ddots & \ddots & \vdots \\ {\bf E}_{1,N} & {\bf E}_{2,N} & \cdots & {\bf E}_{M,N} \end{array}\right] ~\in \mathbb{C}^{NM \times NM} 
\label{P}
\end{align}
 be the row-column interleaver permutation matrix such that ${\bf s}={\bf P}\cdot\thicktilde{\bf x}$ and ${\bf r}={\bf P}\cdot\thicktilde{\bf y}$ where ${\bf E}_{i,j} \in \mathbb{C}^{M \times N}$ is defined as
 \begin{align} {\bf E}_{i,j}(i',j')  = \left. \left\lbrace \begin{array}{ll}  1, & \text{if } i'=i \text{ and } j'=j \\ 0, & \text{otherwise.} \end{array}\right. \right.
     \label{cond_P} \end{align}
Such permutation is known in the literature as a {\em perfect shuffle,} and has the following property \cite{perm}: given square matrices ${\bf A}$ and ${\bf B}$ 
\begin{equation}{\bf A}\otimes{\bf B}={\bf P}\cdot({\bf B}\otimes{\bf A})\cdot{\bf P}^\text{T}.\label{prop1}\end{equation}
The input-output relation in (\ref{time_delay_io}) can now be written as
\begin{align}
    & ({\bf P}^{\rm T}\cdot{\bf r})=\thicktilde{\bf H}\cdot({\bf P}^{\rm T}\cdot{\bf s})+\thicktilde{\bf w}.\label{td1}
\end{align}
Multiplying both sides of (\ref{td1}) on the left by ${\bf P}$, the input-output relation can  be expressed in terms of the time-domain channel matrix ${\bf G}={\bf P}\cdot\thicktilde{\bf H}\cdot{\bf P}^{\rm T}$ as
    \begin{equation}
    {\bf r}={\bf G}\cdot{\bf s}+\thickbar{\bf w}.
     \label{time_domain_io}
\end{equation} 
We note that ${\bf G}$ and $\thicktilde{\bf H}$ are {\em similar} matrices and hence share the same eigenvalues \cite{sim_mat}.  
From (\ref{Htd}) using the perfect shuffle property in (\ref{prop1}), the time domain channel matrix ${\bf G}$ can be related to the delay-Doppler domain channel matrix ${\bf H}$ as 
    \begin{align}
        &{\bf G}=({\bf F}_N^{\dag}\otimes{\bf I}_M)\cdot({\bf P}\cdot{\bf H}\cdot{\bf P}^{\rm T})\cdot({\bf F}_N\otimes{\bf I}_M). \label{perm}
\end{align}

As shown in Fig. \ref{mat4} the null symbols added in the delay-Doppler domain act as interleaved guard bands of length $l_{\max}$ in the time-domain vector ${\bf s}$ and thus help in avoiding interference between the time domain blocks ${\bf r}_n$ for $n=0, \cdots, N-1$. 
This forces ${\bf G}$ to be a block-diagonal matrix. As a result, the large matrix equation in (\ref{time_domain_io}) can be split into $N$ parallel smaller linear matrix equations with the blocks ${\bf G}_0, \cdots, {\bf G}_{N-1} \in \mathbb{C}^{M \times M}$ as the corresponding channel matrices. ${\bf G}_n$ are the diagonal blocks of {\bf G} each with a bandwidth of $l_{\max}+1$. The system equation in (\ref{time_domain_io}) can be split and written as
\begin{equation}
    {\bf r}_n={\bf G}_n\cdot{\bf s}_n+\thickbar{\bf w}_n \quad\text{where }~~ n=0, \cdots, N-1. 
     \label{time_block_io}
\end{equation}
Since ${\bf G}={\bf P}\cdot\thicktilde{\bf H}\cdot{\bf P}^{\rm T}$, the non-zero entries of the  $M \times M$ time domain channel sub-matrices ${\bf G}_n$ are related to the entries of the $N \times N$ delay-time channel sub-matrices $\thicktilde{\bf K}_{m,l}$ and  the time-varying complex channel gain for each delay tap $g^{\rm s}(l,q)$ as
\begin{equation}
    g^{\rm s}(l,q)={\bf G}_n(m,m-l)=\thicktilde{\bf K}_{m,l}(n,n)=\thicktilde{\pmb \nu}_{m,l}(n) \label{td2time}
\end{equation}
where $q=m+nM$, $m \in \{l\leq i < M|l \in \mathcal{L}\}$ and $0\leq n<N$.


\section{Low Complexity Iterative Rake Detector}\label{sec1}

\begin{figure*}
\centering
{\includegraphics[trim=15 15 20 15,clip,height=3.1in,width=6in]{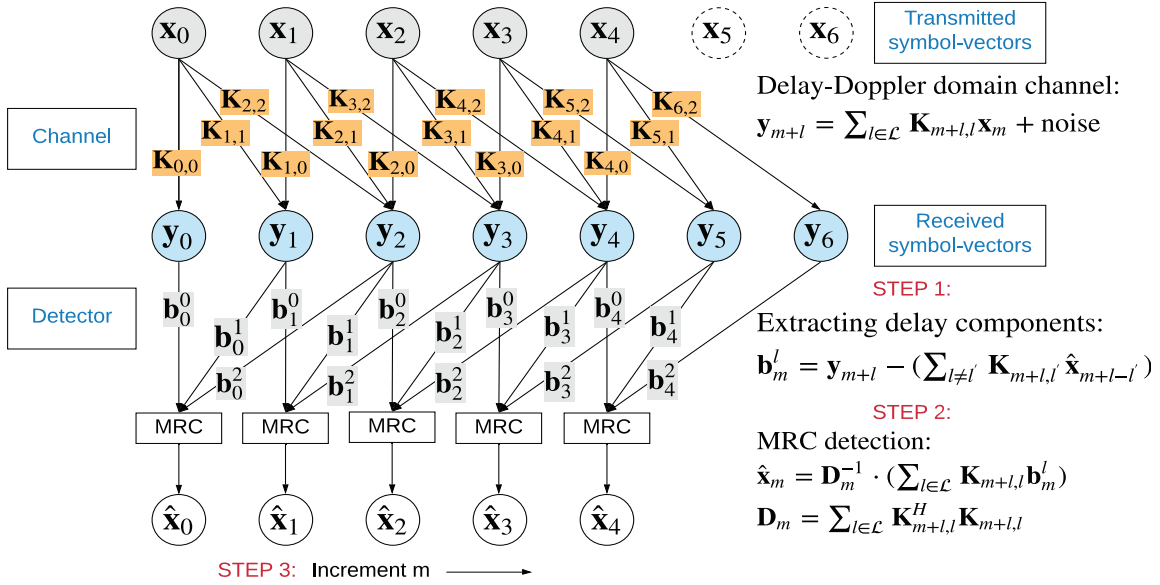}
 \vspace{-4mm}\caption{MRC delay-Doppler domain operation for $M=7$ and the set of discrete delay indices $\mathcal{L}={0,1,2}$. }
\label{Rakeblock}}
\end{figure*}

We can think of the proposed MRC detector as the maximal ratio combining of the channel impaired signal components received at  $L=|\mathcal{L}|\leq P$ different delay branches in the delay-Doppler grid analogous to a CDMA rake receiver as shown in Fig. \ref{Rakeblock}. The noise plus interference (NPI) power in each of these branches is different and depends on the channel response. In each detector iteration, we cancel the estimated inter symbol-vector interference in the branches selected for combining, thereby iteratively improving the post MRC signal to interference plus noise ratio (SINR). 
      
The input output relation between the transmitted and received symbol-vectors ${\bf x}_m$ and ${\bf y}_m$ in (\ref{9}) is given by
\begin{equation}{\bf y}_{m+l}={\sum_{l \in \mathcal{L}}{\bf K}_{m+l,l}\cdot{\bf x}_{m}} + {\bf w}_{m+l} \label{io}\end{equation}
 where ${\bf w}_m$ is iid AWGN noise with variance $\sigma_n^2$. From (\ref{io}), due to the inter-symbol interference caused by delay spread ($l_{\max}T/M$), all symbol-vectors ${\bf x}_{m}$ have a signal component in $L$ received symbol-vectors ${\bf y}_{m+l}$, for $l \in \mathcal{L}$. Let ${\bf b}_m^{l} \in \mathbb{C}^{N \times 1}$ be the channel impaired signal component of ${\bf x}_m$ in the received ${\bf y}_{m+l}$ vector at delay index $m+l$  after removing the interference of the other transmitted symbol-vectors ${\bf x}_{k}$ for $k \neq m$. Assuming we have the estimates of symbol-vectors ${\bf x}_m$ from previous iterations, we can then write ${\bf b}_m^{l}$ for $l\in \mathcal{L}$  as
        \begin{equation}
        {\bf b}_m^{l} ={\bf y}_{m+l}-{\sum_{{{l'}}\in \mathcal{L},{l'}\neq l}{\bf K}_{m+l,{{l'}}}}\cdot\hat{\bf x}_{m+l-{{l'}}}.
        \label{highcomp}\end{equation}
Then from (\ref{io}) and (\ref{highcomp}) for $l\in \mathcal{L}$, we have $L$ equations for the symbol-vector estimates $\hat{\bf x}_{m}$ given as
         \begin{equation}
        {\bf b}_m^{l} ={{\bf K}_{m+l,l}}\cdot{\hat{\bf x}_m}+ {\bf w}_{m+l}+\text{interference}
        \label{dfe}\end{equation}
in the delay branch with index $l$ due to error in the current estimates of the interfering symbol-vectors ${\bf x}_{m+l-{p}}$ for $l \neq {p}$.
In the proposed scheme, instead of estimating the transmitted symbol-vector $\hat{\bf x}_m$ separately from each of the $L$ equations in (\ref{dfe}), we perform maximal ratio combining (\ref{Rm}) of the estimates ${\bf b}_m^{l}$ followed by symbol-by-symbol QAM demapping using (\ref{ML}). 
The vector output of the maximal ratio combiner, ${\bf c}_m \in \mathbb{C}^{N \times 1}$, is given by 
 \begin{equation}{\bf c}_m ={\bf D}_{m}^{-1}\cdot{\bf g}_m \label{Rm} \end{equation}
where
\begin{equation}
    {\bf  D}_{m}={\sum_{l \in \mathcal{L}}{\bf K}_{m+l,l}^{\dag}\cdot{{\bf K}_{m+l,l}}}\label{denom}
\end{equation}
\begin{equation}
    {\bf g}_m=\sum_{l \in \mathcal{L}}{\bf K}_{m+l,l}^{\dag}\cdot{\bf b}_m^{l}\label{num}
\end{equation}
and the hard estimates are given by
\begin{equation} {\bf \hat{x}}_m(n)=\arg \min _{a_j\in \mathcal {Q}} \left |{a_j-{\bf c}_m(n)}\right |. \label{ML}\end{equation}
where $a_j$ is signal from the QAM alphabet $\mathcal{Q}$, with  $j=1,\ldots,|\mathcal{Q}|$ and $n=0,\ldots,N-1$.
Let $\mathcal{D}(.)$ denote the decision on the estimate ${\bf c}_m$ in every iteration such that $\hat{\bf x}_m^{(i)}=\mathcal{D}({\bf c}_m^{(i)})$. Hard-decision function $\mathcal{D}(c)$ is given by the maximum likelihood (ML) criterion in (\ref{ML}). 
\begin{algorithm}
\SetAlgoLined
{\bf Input}: ${\bf H}$, ${\bf D}_m$, ${\bf y}_m$, ${\bf x}_m={\bf 0}_N$\quad$\forall$ $m=0, \ldots, M-1$\\
    \For{i=1:max iterations}{
    \For{$m=0:M'-1$}{
       \For{$l \in \mathcal{L}$}{
            $ {\bf b}_m^{l} ={\bf y}_{m+l}-{\sum_{{p}\neq l}{\bf K}_{m+l,{p}}}\cdot\hat{\bf x}_{m+l-{p}}$\\
    }
    ${\bf g}_m=\sum_{l\in \mathcal{L}}{\bf K}_{m+l,l}^{\dag}\cdot{\bf b}_m^{l}$\\
    ${{\bf c}}_m ={\bf D}_m^{-1}\cdot{\bf g}_m$\\
    ${\bf \hat{x}}_m=\mathcal{D}({\bf c}_m)\quad(\text{or}\quad\hat{\bf x}_m={\bf c}_m$)$^5$\\
    
    }}  
    {\bf Output}: ${\bf \hat{x}}_m$
\label{algo1}\caption{MRC in delay-Doppler domain.}
\end{algorithm}
Once we update the estimate $\hat{\bf x}_m$, we increment $m$ and repeat the same to estimate all  $M'=M-l_{max}$ information symbol-vectors ${\bf \hat{x}}_m$  using the updated estimates\footnote{Alternatively, a soft estimate can also be used in conjunction with an outer coding scheme as described in Section \ref{Turbo_Rake}.} of the previously decoded symbol-vectors in the form of a decision feedback equalizer (DFE) as shown in Fig. \ref{Rakeblock}. Note that the DFE action leads to sequential updates whereas alternatively, using only the previous iteration estimates leads to parallel updates. We verified experimentally that parallel updates result in slower convergence. Algorithm 1 shows the delay-Doppler domain MRC operation (also see Fig. \ref{Rakeblock}).

\subsection{Reduced complexity delay-time domain implementation}
In (\ref{highcomp}), for each symbol-vector ${\bf x}_m$, we need to compute $L$ vectors ${\bf b}_m^{l}$. This operation requires $L(L-1)$ products between matrices ${\bf K}_{m,l}$ and estimated symbol-vectors $\hat{\bf x}_{m-l}$. We can take advantage of the redundant operations to reduce the complexity. Let us define the {\em residual  noise plus interference} (RNPI) term in the $i$-th iteration
\begin{equation}{ \Delta}{\bf y}_m^{(i)}={\bf  y}_{m}-{\sum_{l \in \mathcal{L}}{\bf K}_{m,l}}\cdot\hat{\bf x}_{m-l}^{(i)}\label{deltay}\end{equation} 
which can be considered as the residual error in the reconstructed received delay-Doppler domain symbols due to error in estimation of the transmitted symbols. Note that symbol-vectors $\hat{\bf x}_m$ are estimated in increasing order for $m=0, \ldots, M'-1$. Therefore, for estimating the symbol-vector ${\bf x}_m$, only the symbol-vectors $\hat{\bf x}_{m+p}$, for $p<0$, have updated estimates available in the current iteration. For $p \geq 0$, the previous iteration estimates are used. From (\ref{highcomp}) and (\ref{deltay}), ${\bf b}_m^l$ computation for estimating the symbol-vector ${\bf x}_m$ in the $i$-th iteration can be written as
\begin{equation}{\bf b}_m^{l} ={ \Delta}{\bf y}_{m+l}^{(i)}+{\bf K}_{m+l,l}\cdot\hat{\bf x}_{m}^{(i-1)}.\label{bml}
\end{equation}
Substituting (\ref{bml}) for ${\bf b}_m^{l}$ in  (\ref{num}), the direct computation of ${\bf b}_m^l$ can be avoided by writing ${\bf g}_m^{(i)}$ for the $i$-th iteration as 
\begin{align}
    {\bf g}_m^{(i)}&=\sum_{l \in \mathcal{L}}{\bf K}_{m+l,l}^{\dag}\cdot{ \Delta}{\bf y}^{(i)}_{m+l}+\left(\sum_{l \in \mathcal{L}}{\bf K}_{m+l,l}^{\dag}\cdot {\bf K}_{m+l,l}\right)\cdot \hat{\bf x}_m^{(i-1)} \nonumber\\&
    =\sum_{l \in \mathcal{L}}{\bf K}_{m+l,l}\cdot{ \Delta}{\bf y}^{(i)}_{m+l}+{\bf D}_m\cdot\hat{\bf x}_m^{(i-1)}.\label{gm1}
\end{align}
Then from (\ref{Rm}) and (\ref{gm1}), the MRC output at the $i$-th iteration can be written as
\begin{align}
    {\bf c}_m^{(i)}&=\hat{\bf x}_m^{(i-1)}+{\bf D}_m^{-1}\cdot{\Delta}{\bf g}_m^{(i)} \label{cm1}
\end{align}
where
\begin{equation}{\Delta}{\bf g}_m^{(i)}=\sum_{l \in \mathcal{L}}{\bf K}_{m+l,l}^{\dag}\cdot{\Delta}{\bf y}_{m+l}^{(i)}\label{qm}
\end{equation}
The vector ${\Delta}{\bf g}_m^{(i)}$ in (\ref{qm}) is the maximal ratio combining of the RNPI's in all the delay branches $({\bf y}_{m+l} \text{ for } l \in \mathcal{L})$ having a component of ${\bf x}_m$ in them. 

In the $i$-th iteration, for every estimated symbol-vector ${\bf x}_m$, $L$ RNPI vectors ${\Delta}{\bf y}_{m+l}^{(i)}$ need to be updated. which costs $L^2$ matrix-vector products. However, the complexity of (\ref{deltay}) can be reduced by storing and updating the initial RNPI vectors ${\Delta}{\bf y}_{m}^{(0)}$. The $L$ RNPI vectors which have a component of the most recently estimated symbol-vector are updated as follows,
\begin{equation}
{\Delta}{\bf y}_{m+l}^{(i)}\leftarrow{\Delta}{\bf y}_{m+l}^{(i)}-{\bf K}_{m+l,l}\cdot({{\bf x}}_{m}^{(i)}-{{\bf x}}_{m}^{(i-1)}).\label{deltay2}
\end{equation}
The number of matrix-vector products required to compute ${\Delta}{\bf y}_{m}^{(i)}$ has now been reduced from $L^2$ in (\ref{deltay}) to $L$ in (\ref{deltay2}). Moreover, as described in Section II-E, the matrix-vector products in (\ref{qm}) and (\ref{deltay2}) are products between circulant matrices ${\bf K}_{m,l} \in {\mathbb C}^{N \times N}$ and column vectors ${\bf x}_{m}$ or $\Delta{\bf y}_{m}\in {\mathbb C}^{N \times 1}$ which can be converted to element-wise product of vectors $\thicktilde{\pmb \nu}_{m,l} \circ \thicktilde{\bf x}_{m}$ or $\thicktilde{\pmb \nu}_{m,l} \circ \thicktilde\Delta{\bf y}_{m}$, respectively, in the delay-time domain with a complexity of $N$ complex multiplications. Let the superscript $\sim$ denotes the $N$-IFFT of a vector (i.e., $\thicktilde{\bf a}={\bf F}_N^\text{H}\cdot {\bf a}$). The equations (\ref{cm1}), (\ref{qm}) and (\ref{deltay2}) can now be written in corresponding delay-time domain as
\begin{align}
    \thicktilde{\bf c}_m^{(i)}&=\thicktilde{\bf x}_m^{(i-1)}+{\Delta}\thicktilde{\bf g}_m^{(i)}\oslash\thicktilde{\bf d}_m \label{cmtd}
\end{align}
\begin{equation}{\Delta}\thicktilde{\bf g}_m^{(i)}=\sum_{l \in \mathcal{L}}\thicktilde{\pmb \nu}_{m+l,l}^{\ast}\circ{\Delta}\thicktilde{\bf y}_{m+l}^{(i)}\label{qmtd}\end{equation}
\begin{equation}
{\Delta}\thicktilde{\bf y}_{m+l}^{(i)}\leftarrow{\Delta}\thicktilde{\bf y}_{m+l}^{(i)}-\thicktilde{\pmb \nu}_{m+l,l}\circ({\thicktilde{\bf x}}_{m}^{(i)}-{\thicktilde{\bf x}}_{m}^{(i-1)})\label{deltaytd}
\end{equation}
where 
\begin{equation}\thicktilde{\bf d}_m=\sum_{l \in \mathcal{L}}\thicktilde{\pmb \nu}_{m+l,l}^{\dag}\circ\thicktilde{\pmb \nu}_{m+l,l}\end{equation} 
which can be computed in only $NL$ complex multiplications. 
\begin{algorithm}
\SetAlgoLined
{\bf Input}: $\thicktilde{\bf H}$, $\thicktilde{\bf d}_m$, $\thicktilde{\bf x}_m^{(0)}$, $\thicktilde{\bf y}_m$ $\forall$ $m=0, \ldots, M-1$\\
\For{$m=0:M'-1$}{
        $\Delta\thicktilde{{\bf y}}^{{(0)}}_{m}=\thicktilde{{\bf y}}_m-{\sum_{l \in \mathcal{L}}\thicktilde{\pmb{\nu}}_{m,l}\circ{\thicktilde{\bf x}_{m-l}^{(0)}}}$\\}
    \For{i=1:max iterations}{
    $\Delta\thicktilde{{\bf y}}^{{(i)}}=\Delta\thicktilde{{\bf y}}^{{(i-1)}}$\\
    \For{$m=0:M'-1$}
    {
       ${\Delta}\thicktilde{\bf g}_m^{(i)}=\sum_{l \in \mathcal{L}}\thicktilde{\pmb \nu}_{m+l,l}^{\ast}\circ{\Delta}\thicktilde{\bf y}_{m+l}^{(i)}$\\
    $\thicktilde{{\bf c}}_m^{(i)} =\thicktilde{{\bf x}}_m^{(i-1)}+{\Delta}\thicktilde{\bf g}_m^{(i)}\oslash\thicktilde{\bf d}_m$\\
    $\thicktilde{\bf {x}}_m^{(i)}={\bf F}_N^{\dag}\cdot\mathcal{D}({\bf F}_N\cdot\thicktilde{{\bf c}}_m^{(i)})$ \quad(or\quad$\thicktilde{\bf {x}}_m^{(i)}=\thicktilde{\bf c}_m^{(i)}$)\\
    \For{$l \in \mathcal{L}$}{
    ${\Delta}\thicktilde{\bf y}_{m+l}^{(i)}\leftarrow{\Delta}\thicktilde{\bf y}_{m+l}^{(i)}-\thicktilde{\pmb \nu}_{m+l,l}\circ({\thicktilde{\bf x}}_{m}^{(i)}-{\thicktilde{\bf x}}_{m}^{(i-1)})$\\
    }
    }
    {\bf if} ({$||{\Delta}\thicktilde{\bf y}^{(i)}|| \geq ||{\Delta}\thicktilde{\bf y}^{(i-1)}||$}) {\bf then EXIT} 
    
    }
    {\bf Output}: $\hat{\bf x}_m={\mathcal{D}}({\bf F}_N\cdot\thicktilde{\bf x}_m)$
\caption{Reduced complexity MRC in delay-time domain \label{algo2}.}
\end{algorithm}
\subsubsection{Computational complexity per iteration}
Overall complexity per iteration for calculating ${\Delta}\thicktilde{\bf g}_m^{(i)}$, $\thicktilde{\bf c}_m^{(i)}$ and ${\Delta}\thicktilde{\bf y}_m^{(i)}$ for all symbol-vectors is $M'(2L+1)N$ complex multiplications.
The redundant FFT computations can be avoided by storing the Fourier transform of the $M'L$ Doppler spread vectors ${\pmb \nu}_{m,l}$, the $M'$ initial symbol-vector estimates ${\bf x}_m^{(0)}$ and the RNPI vectors $\Delta\thicktilde{\bf  y}^{(0)}_{m}$ in (\ref{deltay2}). The hard decision estimates require the delay-time vectors to be transformed into the delay-Doppler domain  and back using {\em two} $N$-IFFT operations (which requires $2N\log_2(N)$ complex multiplications) per symbol-vector. Algorithm \ref{algo2} shows the low complexity delay-time domain MRC implementation. The detector iterations are stopped when the overall RNPI error $\Delta\thicktilde{\bf y}=[\Delta\thicktilde{\bf y}_0^\text{T}, \Delta\thicktilde{\bf y}_{1}^\text{T}, \cdots, \Delta\thicktilde{\bf y}_{M-1}^\text{T}]^\text{T}$ due to the estimation error in symbol-vectors stops reducing. 
\subsubsection{Initial computational complexity}
In the proposed detector, the initial computations include generating all the entries of the matrices ${\bf H}$ and $\thicktilde{\bf H}$, which requires computing the vectors ${\pmb \nu}_{m,l}$ and their Fourier transform $\thicktilde{\pmb \nu}_{m,l}$ for all $m=0, \ldots, M'-1$ and $l \in \mathcal{L}$. Assuming the integer delay-Doppler channel parameters $(h_i, k_i, l_i)$ are known for $i=1, 2, \ldots, P$, the channel Doppler spread vectors ${\pmb \nu}_{m,l}$ can be easily computed using the relations given in (\ref{4}) and (\ref{g_lt_to_v_ml6}). 

Let $K_l$ be the number of non-zero channel coefficients in each vector ${\pmb \nu}_{m,l}$ (or paths with different Doppler shift in the same delay bin $l \in \mathcal{L}$) such that total number of channel coefficients or propagation paths as seen by the OTFS receiver is $P=\sum_{l \in \mathcal{L}}K_l$. The number of complex multiplications required to compute the $M'L$ vectors ${\pmb \nu}_{m,l}$ using (\ref{g_lt_to_v_ml6}) is $M'\sum_{l \in \mathcal{L}}K_{l}=M'P$. The OTFS channel matrix ${\bf H}$ (or equivalently the vectors
${\pmb \nu}_{m,l}$) can then be generated in $M'P$ complex multiplications.

For the delay-time domain MRC operation in Algorithm \ref{algo2},
$\thicktilde{\pmb \nu}_{m,l}$ ($N$-IFFT of ${\pmb \nu}_{m,l}$) can be computed in $\min\{Nk_l,N\log_2(N)\}$ complex multiplications, since there are only $K_l$ non-zero channel coefficients in each delay tap $l$. Then, the number of complex multiplications required to compute $\thicktilde{\bf H}$ (or equivalently all the $\thicktilde{\pmb \nu}_{m,l}$) is upper bounded by $M'N \sum_{l}K_l=M'NP$. 

Alternatively, for the fractional Doppler case, the complexity of initial computations remains unaffected for the delay-time domain detector as $\thicktilde{\pmb \nu}_{m,l}$ can be generated directly from the channel gains, delays, and Doppler shifts $(h_i,\kappa_i,\ell_i)$ of the $P$ paths, using (\ref{4}) and (\ref{g_lt_to_v_ml22}) with $M'NP$ complex multiplications.

\subsection{Low complexity initial estimate}
In Algorithm 1 and 2, we initially assume that all the $\mathcal{Q}$-QAM signals $a_j$ are equally likely and the mean of $a_j$'s is zero and so we initialize $\hat{{\bf x}}_m^{(0)} ={\bf 0}_N$, for all $m$. The MRC detector complexity per iteration is of the order $O(NML)$ and the overall complexity scales linearly with the number of iterations. 

However, a better initial estimate of the OTFS symbols instead of $\hat{\bf x}_m={\bf 0}_N$ may reduce the required number of MRC iterations and to reach convergence. Assuming ideal pulse shaping waveform, a single tap equalizer in the time-frequency domain can provide an improved low complexity initial estimate.

Following the remark in Section \ref{Sec:IOrelationDD} and \cite{WCNC_paper}, we define ${\bf H}_{\text{dd}} \in \mathbb{C}^{M \times N}$, the delay-Doppler domain channel impulse response matrix for the ideal pulse shaping waveform case, 
\begin{align} {\bf H}_{\text{dd}}(m,n) = \left. \left\lbrace \begin{array}{ll} \nu_{l}(\kappa), & \text{if } m=l, n=[\kappa]_N\\0, & \text{otherwise.} \end{array}\right. \right. \end{align}

For the fractional Doppler case (when $\kappa$ is a real number). the ideal channel response can be written in terms of the Doppler spread vectors as ${\bf H}_{\text{dd}}=[{\pmb \nu}_{0,0}, {\pmb \nu}_{1,1}, \cdots, {\pmb \nu}_{M-1,M-1}]^\text{T}$. The corresponding time-frequency channel response for the ideal pulse shaping waveform is obtained by an inverse symplectic finite fourier transform (ISFFT) operation on the delay-Doppler channel as
\begin{align}{\bf H}_{\text{tf}}&={\bf F}_M\cdot{\bf H}_{\text{dd}}\cdot{\bf F}^{\text{H}}_N\label {Htf} \\
&={\bf F}_M\cdot[{\pmb \nu}_{0,0}, {\pmb \nu}_{1,1}, \cdots, {\pmb \nu}_{M-1,M-1}]^\text{T}\cdot{\bf F}^{\text{H}}_N \nonumber\\
&={\bf F}_M\cdot[\thicktilde{\pmb \nu}_{0,0}, \thicktilde{\pmb \nu}_{1,1}, \cdots, \thicktilde{\pmb \nu}_{M-1,M-1}]^\text{T}.
\label {Htf2}\end{align}
Similarly, the received time-frequency samples can be obtained by the ISFFT operation on the received delay-Doppler domain samples as
\begin{align}{\bf Y}_{\text{tf}}={\bf F}_M\cdot{\bf Y}\cdot{\bf F}^{\text{H}}_N={\bf F}_M\cdot[\thicktilde{\bf y}_0, \thicktilde{\bf y}_1, \cdots \thicktilde{\bf y}_{M-1}]^\text{T}.\label{ytf}\end{align}
Since in the ideal pulse shaping waveform case, circular convolution of the channel and transmitted symbols in the delay-Doppler domain transforms to element-wise product in the time-frequency domain, we estimate the transmitted samples in the time-frequency domain by a single tap  minimum mean square error (MMSE) equalizer 
\begin{equation}
    {\bf \hat{X}}_{\text{\text{tf}}}(m,n)=\frac{{\bf H}_{\text{tf}}^{\ast}(m,n)\cdot{\bf Y}_{\text{tf}}(m,n)}{\lvert {\bf H}_{\text{tf}}(m,n)\rvert^{2}+\sigma_w^2}
    \label{20}
\end{equation}
for $m=0,\ldots,M-1$ and $n=0,\ldots,N-1$.

The time-delay domain initial estimates of the OTFS symbol-vectors can then be obtained by the Heisenberg transform operation on the time-frequency domain estimates as 
    \begin{equation}
     [\thicktilde{\bf x}_0^{(0)}, \thicktilde{\bf x}_1^{(0)}, \cdots \thicktilde{\bf x}_{M-1}^{(0)}]^\text{T}={\bf F}^{\dag}_M\cdot{\bf \hat{X}}_{\text{tf}}. \label{xx}
    \end{equation}
Note that $\thickbar{\pmb \nu}_{m,l}={\bf 0}_N$ for $l \not\in \mathcal{L}$ and hence the operation in (\ref{Htf2}) can be computed in $\min\{NML,NM\log_2(M)\}$ complex multiplications. Since we have already computed $\thicktilde{\pmb \nu}_{m,l}$, and $\thicktilde{\bf y}$ is just a shuffled version of the received time-domain samples, the overall number of computations (for the steps in (\ref{Htf2}), (\ref{ytf}), (\ref{20}) and (\ref{xx})) required for the initial estimate is upper bounded by  $NM(L+2\log_2(M)+3)$, which is comparable to the complexity of one detector iteration $NM'(2L+1)$.

\subsection{Condition for Detector Convergence}

{In this section, we cast the delay-time algorithm (Algorithm \ref{algo2}) in the time-domain with the purpose of analysing the detector convergence using the properties of Jacobi and Gauss Seidel iterative methods for solving linear equations \cite{LSBook,GSBook}. The basic principle of iterative MRC operation in the delay-time domain with sequential updates given in  (\ref{cmtd})-(\ref{deltaytd}) can be compactly expressed as
\begin{equation}
    \thicktilde{\bf x}^{(i)}=\thicktilde{\bf x}^{(i-1)}+\thicktilde{\bf D}^{-1}\thicktilde{\bf H}^{\dag}(\thicktilde{\bf y}-\thicktilde{\bf H}\thicktilde{\bf x}^{(i-1)})\label{tdalgo}
\end{equation}
when using parallel updates (i.e. without DFE),
where $\thicktilde{\bf D}$ is the matrix containing diagonal elements of $\thicktilde{\bf H}^{\dag}\thicktilde{\bf H}$.
The rows and columns of the delay-time channel matrix $\thicktilde{\bf H}$ are perfectly shuffled using the permutation matrix ${\bf P}$ to obtain a {\em similar}, block diagonal time-domain channel matrix ${\bf G}$ as explained in Section II-F. This allows the equivalent operation in (\ref{tdalgo}) to be split and executed in parallel for each independent time domain block ${\bf G}_n$ as
\begin{equation}
    {\bf s}_n^{(i)}={\bf s}_n^{(i-1)}+{\bf D}_n^{-1}{\bf G}_n^{\dag}({\bf r}_n-{\bf G}_n{\bf s}_n^{(i-1)})\label{tdalgo1}
\end{equation}
where ${\bf D}_n$ is the matrix containing the diagonal elements of ${\bf G}_n^{\dag}{\bf G}_n$.
Equation (\ref{tdalgo1}) can be written in the form
\begin{align}
    &{\bf s}_n^{(i)}=-{\bf T}^{\rm J}_n\cdot{\bf s}_n^{(i-1)}+{\bf Q}^{\rm J}_n\cdot{\bf z}_n \nonumber \\
&{\bf T}^{\rm J}_n={\bf D}_n^{-1}\cdot({\bf L}_n+{\bf L}_n^{\dag}),\quad{\bf Q}^{\rm J}_n={\bf D}_n^{-1},\quad{\bf z}_n={\bf G}_n^{\dag}{\bf r}_n \label{sys_eq}
\end{align}
where ${\bf L}_n$ and ${\bf L}_n^{\dag}$ are the matrices containing the strictly lower and upper triangular parts of the Hermitian matrix ${\bf R}_n={\bf G}_n^{\dag}{\bf G}_n$. Finally, we observe that the parallel update formulation in (\ref{sys_eq}) matches the classic Jacobi iterative method (hence the superscript 'J' in ${\bf T}_n^{\rm J}$) for solving linear equations, \cite{LSBook}. 

We now focus on the sequential update method given in Algorithm 1 and 2 based on the DFE operation. Note that, in Algorithm \ref{algo2}, the linear matrix equation in (\ref{tdalgo}) is solved block-wise with low complexity, where the latest estimates of the symbol-vectors calculated in the current iteration are used in estimating the next symbol-vector as in a DFE
\begin{equation}
{\bf s}_n^{(i)}={\bf s}_n^{(i-1)}+{\bf D}_n^{-1}({\bf z}_n-\underbrace{{\bf L}_n{\bf s}_n^{(i)}}_{(a)}-\underbrace{{\bf L}_n^{\dag}{\bf s}_n^{(i-1)}}_{(b)})\label{tdalgo3}
\end{equation}
where $(a)$ and $(b)$ denote the contribution of the current and previous-iteration estimates, respectively. We can modify (\ref{sys_eq}) for the DFE iterative method in (\ref{tdalgo3}) as 
\begin{align}
    &{\bf s}_n^{(i)}=-{\bf T}^{\rm GS}_n\cdot{\bf s}_n^{(i-1)}+{\bf Q}^{\rm GS}_n\cdot{\bf z}_n \nonumber \\
&{\bf T}^{\rm GS}_n=({\bf D}_n+{\bf L}_n)^{-1}\cdot{\bf L}_n^{\dag},\quad{\bf Q}^{\rm GS}_n=({\bf D}_n+{\bf L}_n)^{-1} \label{Tn}
\end{align}
and observe that Algorithm \ref{algo2} coincides with the well studied Gauss Seidel (GS) method available in the literature \cite{LSBook,GSBook}. Algorithm \ref{algo3} shows the equivalent time domain GS method implementing Algorithm \ref{algo2}.
\begin{algorithm}
\SetAlgoLined
{\bf Input}: ${\bf r}$, ${\bf G}$\\
\For{$n=0:N-1$}{
${\bf R}_n={\bf G}_n^{\dag}\cdot{\bf G}_n$\\
${\bf z}_n={\bf G}_n^{\dag}\cdot{\bf r}_n$\\
${\bf L}_n=\text{strictly lower triangular part}\{{\bf R}_n$\}\\
${\bf T}^{\rm GS}_n=({\bf D}_n+{\bf L}_n)^{-1}\cdot{\bf L}_n^{\dag}$\\
${\bf Q}^{\rm GS}_n=({\bf D}_n+{\bf L}_n)^{-1}$\\
}
$\hat{\bf s}^{(0)}={\bf P}\cdot({\bf I}_M\otimes{\bf F}_N^{\dag})\cdot {\hat{\bf x}}^{(0)}$\\
\For{$i=1$:max iterations}{
\For{$n=0:N-1$}{
$\hat{\bf s}_n^{(i)}=-{\bf T}^{\rm GS}_n\cdot\hat{\bf s}_n^{(i-1)}+{\bf Q}^{\rm GS}_n\cdot{\bf z}_n$\\
}
{\bf if} ({$||{\bf r}-{\bf G}\cdot\hat{\bf s}^{(i)}|| \geq ||{\bf r}-{\bf G}\cdot\hat{\bf s}^{(i-1)}||$}) {\bf then EXIT}
}
{\bf Output}: $\hat{\bf x}=({\bf I}_M\otimes{\bf F}_N)\cdot({\bf P}\cdot {\hat{\bf s}^{(i)}})$
\caption{MRC delay-time domain operation principle in the form of time domain Gauss-Seidel method \label{algo3}.}
\end{algorithm}

Both Jacobi and GS methods are used to iteratively find the least squares solution
\begin{equation}
   \hat{\bf s}_n=\min_{\hat{\bf s}_n} ||{\bf z}_n-{\bf R}_n\hat{\bf s}_n||^2
\end{equation}
of the $M$-dimensional linear system of equations
\begin{equation}
    {\bf z}_n={\bf R}_n\cdot{\bf s}_n+\thickbar{\bf w}_n
    \label{MRC_matrix}
\end{equation} 
 where ${\bf R}_n \in \mathbb{C}^{M \times M}$ and $\hat{\bf s}_n, {\bf z}_n \in \mathbb{C}^{M \times 1}$. We further assume that the time-domain correlation matrix ${\bf R}_n={\bf G}_n^{\dag}{\bf G}_n$ is non-singular and hence positive definite Hermitian. 

In  \cite{LSBook,GSBook}, it is shown that the iteration method (\ref{sys_eq}) for the linear system in (\ref{MRC_matrix}) is convergent, if $\rho({\bf T}^{\rm GS}_n)<1$, where $\rho({\bf T}^{\rm GS}_n)$ is the spectral radius\footnote{Spectral radius of a matrix is the largest absolute value of its eigenvalues.} of the square matrix ${\bf T}^{\rm GS}_n$ \cite{LSBook,GSBook}. For the Jacobi method, $\rho({\bf T}^{\rm J}_n)<1$ if ${\bf R}_n$ is diagonally dominant, which depends on the channel and cannot be guaranteed. However, the GS method is known to converge faster and convergence is guaranteed under more general conditions than the Jacobi method \cite{LSBook,GSBook}. In Appendix we prove the following lemma
\begin{lemma} \label{rho}
The GS iterative method for the solution of (\ref{MRC_matrix}) is converging  (i.e., $\rho({\bf T}^{\rm GS}_n)<1$) if ${\bf R}_n$ is a  positive definite Hermitian matrix. Furthermore, $\rho({\bf T}^{\rm GS}_n)=1$ if ${\bf R}_n$ is a positive semi-definite Hermitian matrix. 
\end{lemma}

We note that the algorithm may still converge even for some channels that result in a positive semi-definite Hermitian matrix ${\bf R}_n$ (i.e., $\rho{({\bf T}^{\rm GS}_n)} = 1$), but this is not guaranteed.

Even though the implementation of the iterative MRC detector in Algorithm \ref{algo3} looks simpler than the one in Algorithm \ref{algo2}, the complexity of initial computations for directly calculating ${\bf R}_n$, ${\bf T}^{\rm GS}_n$ and ${\bf Q}^{\rm GS}_n$ is $O(NML^2)$ complex multiplications since ${\bf G}_n$ is a banded matrix with $L$ non-zero elements in each row. However, in Algorithm \ref{algo2}, the circulant property of the blocks of the channel matrix ${\bf H}$ (due to the placement of null symbols in the OTFS grid as shown in Fig. \ref{mat2}) is utilized to reduce the overall complexity of the initial computations to $O(NML)$ complex multiplications as explained in Section III-A.}
\vspace{-3mm}

\section{Further Improvements}
\subsection{Successive Over Relaxed (SOR) Iterative Rake Detector}
\begin{figure}
    \centering
    {\includegraphics[trim=10 0 0 10,clip,height=2.2in,width=3.4in]{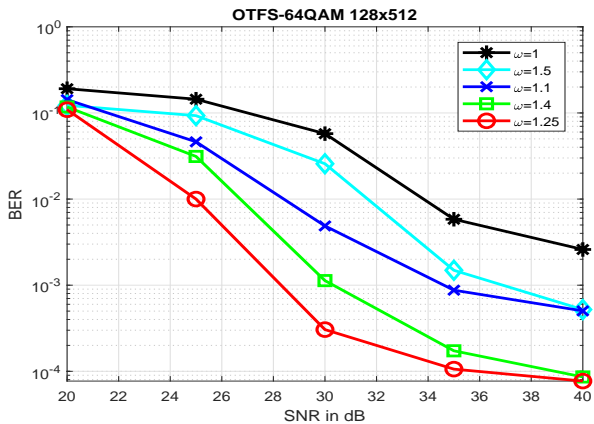}
     \vspace{-3mm}\caption{64-QAM BER performance for different relaxation parameters $\omega$.}
    \label{damp64Q}}
    \end{figure} 
In time domain, the proposed iterative Rake detector is similar to doing $N$ parallel GS iterations on the matched filtered received waveform, as shown in Section III-C. GS and its variants such as successive over-relaxation (SOR) method are well presented in \cite{mathJourn, LSBook, GSBook}. The SOR method is obtained by introducing a relaxation parameter $\omega$ in the GS method (\ref{tdalgo3}) as, 
 \begin{equation}
    {\bf s}_n^{(i)}={\bf s}_n^{(i-1)}+\omega{\bf D}_n^{-1}({\bf z}_n-{\bf L}_n{\bf s}_n^{(i)}-{\bf L}_n^{\dag}{\bf s}_n^{(i-1)}).\label{tdalgo4}
\end{equation}
 The corresponding GS iteration matrix ${\bf T}^{\rm GS}_n$ and ${\bf Q}^{\rm GS}_n$ in Algorithm \ref{algo3} can be modified as
\begin{equation}
    {\bf T}_n^{\omega}=({\bf D}_n+\omega{\bf L}_n)^{-1}\cdot((\omega-1){\bf D}_n+\omega{\bf L}_n^{\dag})
    \label{new_tn}
\end{equation}
\begin{equation}
    {\bf Q}_n^{\omega}=({\bf D}_n+\omega{\bf L}_n)^{-1}.
\end{equation}
In Appendix we prove the following lemma.
\begin{lemma} \label{omega_tn}
The SOR GS iterative method for the solution of (\ref{MRC_matrix}) is converging  (i.e., $\rho({\bf T}^{\omega}_n)<1$) if ${\bf R}_n$ is a  positive definite Hermitian matrix and $0<\omega<2$. 
\end{lemma}
We can then simply modify the proposed delay-time detector Algorithm \ref{algo2} by rewriting (\ref{cmtd}) as
\begin{equation}
\thicktilde{{\bf c}}_m^{(i)} =\thicktilde{{\bf c}}_m^{(i-1)}+\omega({\Delta}\thicktilde{\bf q}_m^{(i-1)}\oslash\thicktilde{\bf d}_m). \label{cmSOR}
\end{equation} 
Note that when $\omega=1$, (\ref{cmSOR}) coincides with (\ref{cmtd}). The relaxation parameter when $\omega>1$ is called the over-relaxation parameter and when $\omega<1$ is called the under relaxation parameter. The computation of the optimal SOR parameter $\omega=\omega_{\text{opt}}$ which minimizes the spectral radius $\rho({\bf T}_n^{\omega})$  requires computing the eigenvalues of the iteration matrix ${\bf T}_n^{\omega}$ , \cite{GSBook,LSBook}.

 The aim is to find the range of values of $\omega$ for which the SOR method converges (see Lemma 2), the set of which denotes the region of convergence, and, if possible, the best value $\omega_{\rm opt}$. The optimum SOR parameter can be analytically calculated given the spectral radius of the Jacobi matrix $\rho({\bf T}^{\rm J}_n)<1$ \cite{mathJourn}. However, it is known that $\rho({\bf T}^{\rm J}_n)<1$ only if ${\bf R}_n$ is diagonally dominant, but this is not guaranteed for all channels. In such cases, the numerical calculation of $\omega_{opt}$  is not practical for large system matrices, rather a region of good performance, within the region of convergence, is easier to find, as suggested by \cite{mathJourn}. Further, when the power delay profile statistical model of the channel is given, the good region for the SOR parameter can be optimized offline by simulation.

\begin{figure}
    \centering
    {\includegraphics[trim=10 0 0 6,clip,height=2.2in,width=3.4in]{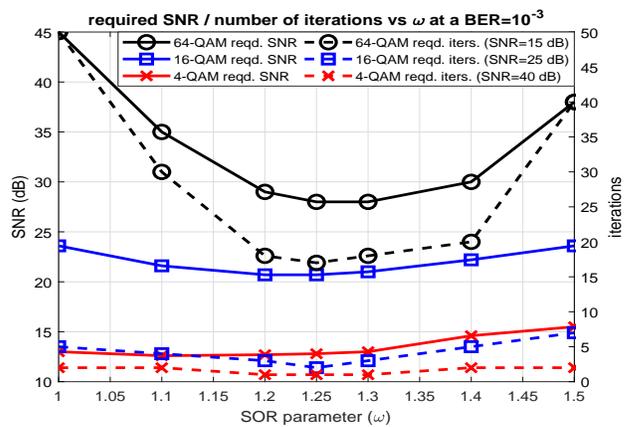}
     \vspace{-3mm}\caption{Error performance and convergence speed of different relaxation parameters $\omega$ for different modulation sizes $|\mathbb{Q}|$ at BER $10^{-3}$.}
    \label{ber_16qam_SOR}}
\end{figure}

In this paper, we try to analyse the effect of $\omega$ and the range of values of good performance by simulation. Fig. \ref{damp64Q} show the BER plot for 64-QAM for different values of $\omega$. In Fig. \ref{ber_16qam_SOR}, we plot the required (abbreviated as reqd. in the plot legend) SNR (labelled as 'Q-QAM reqd. SNR') on the left y-axis alongside the required number of iterations (labelled as 'Q-QAM reqd. iters') on the right y-axis, to achieve a BER of $10^{-3}$ for different modulation sizes, respectively, for different values of $\omega \in [1,1.5]$. The y-axis of the plot represents the SNR (dB) or the iterations depending on the corresponding curve. The maximum number of iterations is set to 50. It can be seen that the optimum $\omega$ for the standard extended vehicular A (EVA) \footnote{The EVA channel power-delay profile (with a maximum speed = 120 km/hr) is given by [0, -1.5, -1.4, -3.6, -0.6, -9.1, -7.0, -12.0, -16.9] dB with excess delay taps $\mathcal{L}'=\mathcal{L}=\{0, 1, 2, 3, 4, 5, 8, 13, 19\}$ normalized to the delay resolution $1/(M\Delta f)$ of an OTFS grid with bandwidth $M\Delta f$, where $M=512$ and $\Delta f=15$ kHz.} channel model \cite{EVA} consistently lies in the interval $[1.2,1.3]$. We can observe that there is a 2.5 dB and 17dB gain at a BER of $10^{-3}$ for 16-QAM and 64-QAM, respectively, due to just the over-relaxation parameter with almost no extra computational complexity. 
The effect of the SOR parameter on the convergence speed of the MRC detector can be seen in Fig. \ref{ber_16qam_SOR} (right y-axis). It shows the number of iterations required to achieve a BER of $10^{-3}$ for different modulation sizes at the corresponding SNR values as given in the plot legend. It can be seen that the biggest reduction in complexity comes at 64-QAM where, the number of iterations required is significantly reduced (by almost 3 times) as compared to the case when SOR parameter $\omega=1$. For 4-QAM and 16-QAM, the optimum SOR parameter approximately halves the number of required iterations.  

 Finally, if no prior knowledge of the channel statistical model is available, we observed by simulation that some performance improvement can still be achieved by setting the value of $\omega$ to slightly above 1. The optimization of $\omega$ with low complexity, for different SNR, channel profiles and number of multipaths will be investigated in future work.



\subsection{Iterative Rake Turbo Decoder \label{Turbo_Rake}}
\begin{figure}
\centering
{\includegraphics[trim=0 0 0 0,clip,height=2.2in,width=3.4in]{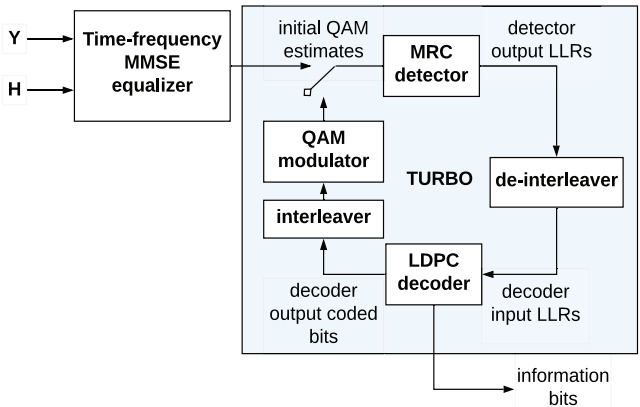}
 \vspace{-3mm}\caption{ OTFS iterative rake turbo decoder operation.}
\label{turbo}}
\end{figure}
 In order to improve FER performance, the turbo decoder principle shown in Fig. \ref{turbo} is proposed. The encoded bits are random interleaved in the frame so as to enhance the delay-Doppler diversity. 

The detector output bit log likelihood ratios (LLR) after random de-interleaving is fed to the low-density parity check (LDPC) decoder. The hard decision coded bits from the LDPC decoder after interleaving and QAM modulation is then fed back to the MRC detector as the input symbol-vector estimates and the process repeats. Overall, one turbo iteration involves one iteration of MRC detector, de-interleaver, LDPC decoder, interleaver, and the QAM modulator. As shown in Fig. \ref{turbo}, for the first iteration, the initial estimate of the QAM symbols is provided by the low complexity MMSE equalizer as explained in Section III-B, after which the initial estimate comes form the LDPC decoder.

From (\ref{cm1}), the soft estimate of the delay-Doppler domain symbol-vector ${\bf c}_m$ after MRC combining can be written as
\begin{equation}
{\bf c}_{m}= {\bf x}_m+{\bf e}_{m} \quad\quad m=0, \ldots M'-1
\label{cmstat}\end{equation}
where ${\bf x}_m$ is the transmitted symbol-vector at delay index $m$ and ${\bf e}_m$ denotes the normalized post MRC NPI vector. We assume that ${\bf e}_m$ follows a zero mean Gaussian distribution with variance ${\sigma}_m^2$. This assumption becomes more accurate as the number of interfering terms increases. Then, the LLR $L_{m,n,b}^{(i)}$ of bit $b$ of the $n$-th transmitted symbol in the estimated symbol-vector ${\bf c}^{(i)}_m$ in the $i$-th iteration can be obtained by 
\begin{align}
    L_{m,n,b}^{(i)}&=\log\left(\frac{Pr(b=0|{\bf c}^{(i)}_m(n))}{Pr(b=1|{\bf c}^{(i)}_m(n))}\right) \nonumber \\&=\log\left(\frac{\sum_{q \in Q_0} \exp({-|{\bf c}^{(i)}_m(n)-q|^2/\sigma_m^2})}{\sum_{q' \in Q_1} \exp({{-|{\bf c}^{(i)}_m(n)-q'|^2}/\sigma_m^2})}\right)
\end{align}
where $\mathcal{Q}_0$ and $\mathcal{Q}_1$ are the subsets of QAM symbols, where the $b$-th bit of the symbol is 0 and 1, respectively. The complexity of LLR calculation can be reduced by the max-log approximated LLR obtained as
\begin{align}
\thicktilde{L}^{(i)}_{m,n,b} = \frac{1}{\sigma^{2}_m} \left(\min_{q\in{\mathcal Q}_{0}} \left\vert{{{{\bf c}^{(i)}_m(n)}{}}}-q\right\vert^2 \!-\! \min_{q'\in{\mathcal Q}_{1}}\left\vert{{{\bf c}^{(i)}_m(n)}{}}-q'\right\vert^2\right).\label{bitllr}
\end{align}
 

In order to compute the bit LLRs, an estimate of the post MRC NPI variance $\sigma_m^2$ is required. Accurate estimation of $\sigma_m^2$ is not straightforward and requires knowledge of the correlation between all the estimated symbol-vectors and RNPI vectors which changes every iteration as well. Since the entries of channel Doppler spread vectors ${\pmb \nu}_{m,l}$ can be assumed to be zero mean, i.i.d. and normal distributed \cite{EVA}, the channel Doppler spread for different delay taps can be assumed to be uncorrelated. i.e., $E[{\pmb \nu}_{m,l}^{\dag}\cdot {\pmb \nu}_{m',{p}}]=0$ for $l \neq {p}$. Furthermore, for the purpose of a simple estimate of the post MRC NPI variance, we assume that RNPI ${\Delta}{\bf y}_m^{(i)}$ in the different delay branches are uncorrelated (i.e., $E[{{\Delta}{\bf y}_m^{\dag}\cdot {\Delta}{\bf y}_p}]=0$ for $m \neq p$ in all iterations) and follows Gaussian distribution. The covariance matrix of the delay-time RNPI vector $\Delta\thicktilde{\bf y}_m$ in the $i$-th iteration  
\begin{align}
{\pmb C}_{m}^{(i)}(j,k)=&({\Delta}\thicktilde{\bf y}_{m}^{(i)}(j)-E\{{\Delta}\thicktilde{\bf y}_{m}^{(i)}\})({\Delta}\thicktilde{\bf y}_{m}^{(i)}(k)-E\{{\Delta}\thicktilde{\bf y}_{m}^{(i)}\})^{\ast}
\end{align}
for $j,k=0, \ldots, N-1$ and $E\{{\Delta}\thicktilde{\bf y}_{m}^{(i)}\}=\frac{1}{N}\sum_{n=1}^{N}{\Delta}\thicktilde{\bf y}_{m}^{(i)}(n)$. Since Fourier transformation is a unitary transformation, the NPI variance remains the same in both domains, and we approximate the post MRC NPI variance for the symbol-vector soft estimate ${\bf c}_m^{(i)}$ in the $i$-th iteration as
\begin{equation}
 \sigma_m^{2{(i)}}=\mathrm{Var}(\thicktilde{\bf e}_m^{(i)})\approx\frac{1}{N}\sum_{l \in L} {\eta}_{m,l}{\mathrm{tr}}({\pmb C}^{(i)}_{m+l}) \label{var_e_m} 
\end{equation}
where ${\eta}_{m,l}=||\thicktilde{\pmb \nu}_{m+l,l}\oslash \thicktilde{\bf d}_m||^2$ is the normalized post MRC channel power in the different delay branches selected for combining. The bit LLR calculation in (\ref{bitllr}) and NPI variance calculation in (\ref{var_e_m}) has a complexity of $2NM\log_2(|\mathcal{Q}|)$ and $NML$, respectively. The LDPC decoder complexity is of the order $C_{\rm LDPC}=O\left(\log_2(|\mathcal{Q}|)NM\right)$. The overall complexity of detection increases by $C_{\rm LDPC}+NM(2\log_2(|\mathcal{Q}|)+L)+$ for every turbo iteration.
    
\begin{figure}
    \centering
    {\includegraphics[trim=10 0 0 10,clip,height=2.2in,width=3.4in]{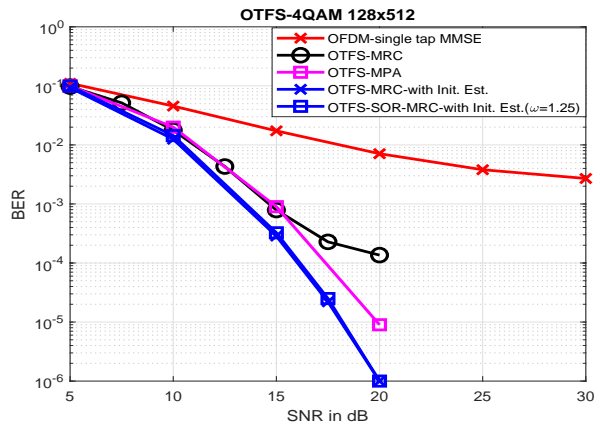}
     \vspace{-3mm}\caption{Uncoded 4-QAM BER Plot : MRC vs MPA vs MMSE-OFDM.}
    \label{ber_4qam}}
    \end{figure}
   \begin{figure}
    \centering
    {\includegraphics[trim=10 0 0 10,clip,height=2.2in,width=3.4in]{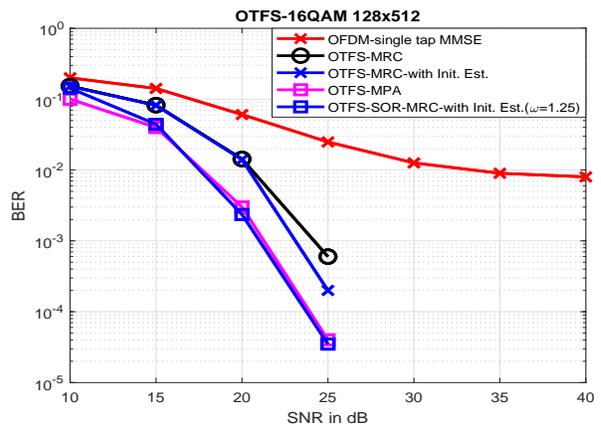}
     \vspace{-3mm}\caption{Uncoded 16-QAM BER Plot : MRC vs MPA vs MMSE-OFDM.}
    \label{ber_16qam}}
    \end{figure}
\section{Simulation Results and Discussion}
For simulations we generate OTFS frames for $N=128$ and $M=512$. The sub-carrier spacing $\Delta f$ is taken as 15 kHz. The maximum delay spread (in terms of integer taps) is taken to be 32 ($l_{\max}=31$) which is approximately 4 ${\mu}s$. The channel delay model is generated according to the standard EVA model (with a speed of 120 km/h) with the Doppler shift for the $i$-th path generated from a uniform distribution $U(0,\nu_{max})$, where $\nu_{max}$ is the maximum Doppler shift \cite{EVA}. We consider one Doppler shifted path per delay tap with $L=9$ and $k_{max}=16$. For our simulations, we assume perfect knowledge of the channel state information at the receiver (see \cite{Ravi3} for practical channel estimation in OTFS). For BER plots, $10^{5}$ frames are send for every point in the BER curve and for FER plots, all simulations run for a minimum of $10^{5}$ frames or until 100 OTFS frame errors are encountered. BER is plotted to show uncoded performance, while FER is used when an outer coding scheme is applied.
 \begin{figure}
    \centering
    {\includegraphics[trim=10 0 0 10,clip,height=2.2in,width=3.4in]{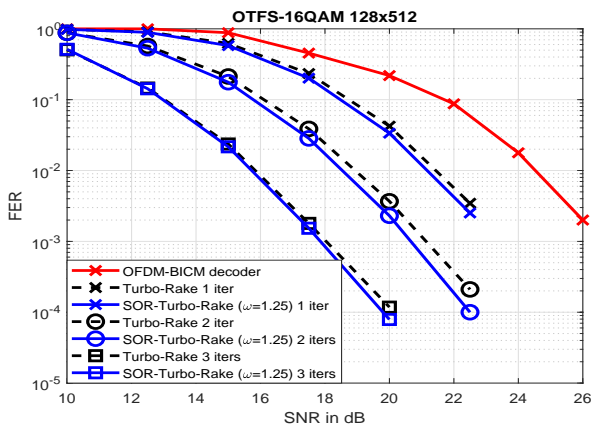}
    \vspace{-3mm}\caption{Turbo 16-QAM FER Plot: MRC vs BIC-MMSE-OFDM.}
    \label{turboplot}}
    \end{figure}
    \begin{figure}
    \centering
    {\includegraphics[trim=10 0 0 10,clip,height=2.2in,width=3.4in]{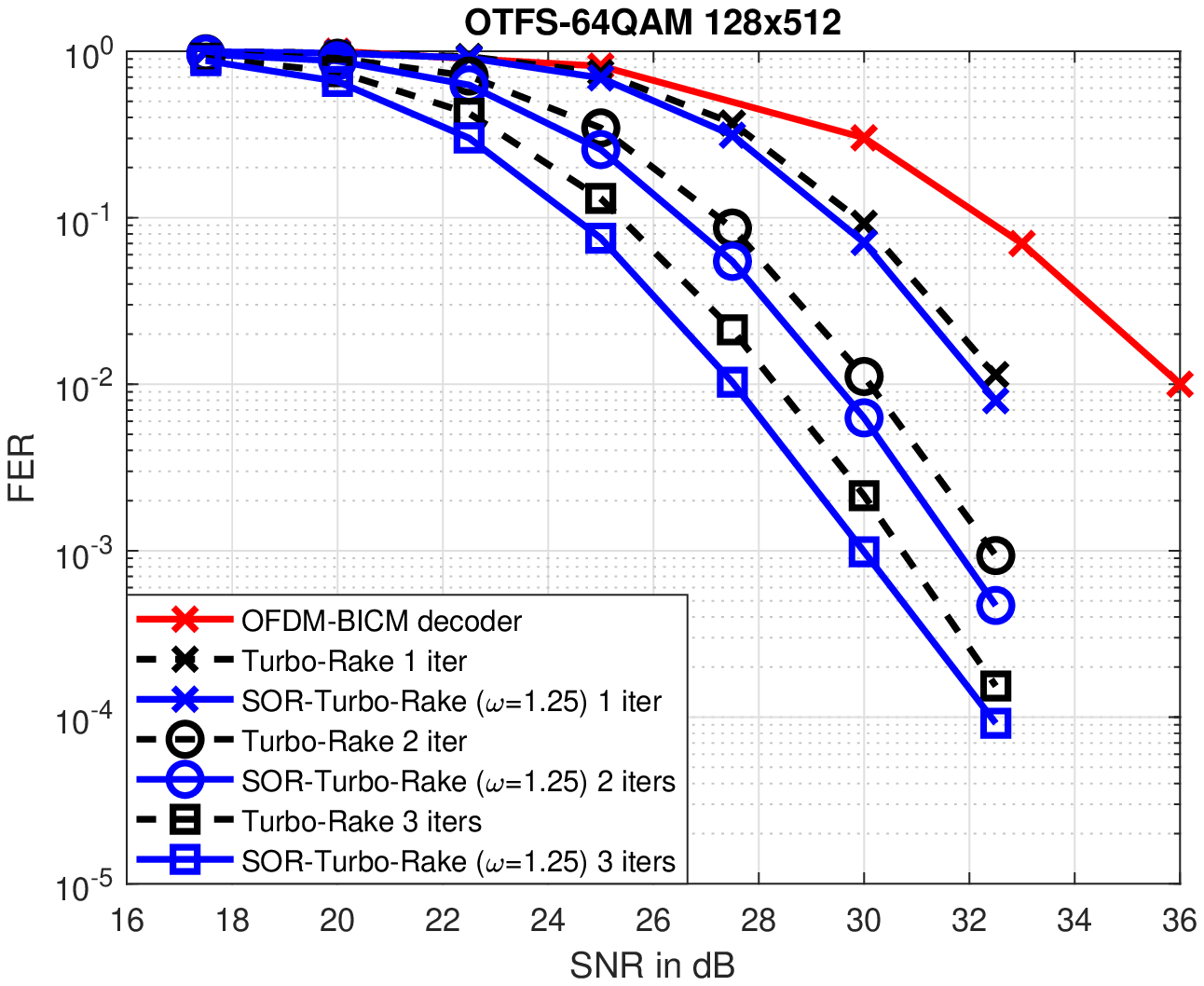}
     \vspace{-3mm}\caption{Turbo 64-QAM FER Plot: MRC vs BIC-MMSE-OFDM. }
    \label{fer_64qam}}
    \end{figure}  
    
Fig. \ref{ber_4qam} shows the BER plot for the MRC detector, with and without the initial estimate in Section III-B, for 4-QAM modulated OTFS waveform with a maximum of 10 iterations\footnote{Iterations are stopped according to the residual NPI convergence criteria in Algorithm \ref{algo2}.}. Performance is compared with the state of the art {\em message passing algorithm} (MPA) described in \cite{Ravi,Ravi1} (labeled as OTFS-MPA in Fig. \ref{ber_4qam} and \ref{ber_16qam}) with a maximum of 10 iterations\footnote{The MPA stopping criteria is based on the convergence of the estimated symbol probabilities \cite{Ravi}.} and the OFDM single tap MMSE equalizer. It can be seen that with the initial estimate (labeled as OTFS-MRC with Init. Est.\footnote{Init. Est. refers to detection with the Initial Estimate in Section III-B.}), there is a $\approx$1 dB gain over the MPA algorithm at a BER of $10^{-3}$. This gain is contributed by the improved SNR due to the MRC operation (or matched-filtering) at the receiver and the initial time-frequency MMSE estimate, which is more reliable for lower modulation sizes like BPSK and 4-QAM, thereby increasing the convergence speed (due to the initial estimates begin closer to the solution). 

Note that the same initial estimates could also be used to improve the performance of MPA. However, the estimates need to be transformed into the delay-Doppler domain and $\mathcal{Q}$-QAM alphabet probabilities for all the information symbols need to be calculated. This would incur a high complexity just to get the improved initial estimate. Moreover, similar to MRC detection, MPA can also be applied on the matched-filtered system matrix ${\bf H}^{\dag}{\bf H}$ instead of ${\bf H}$, but this approximately doubles the MPA complexity, which scales linearly with the number of non-zero elements in the matrix. \cite{Ravi,Ravi1}.

Fig. \ref{ber_16qam} shows the BER plot for the MRC detector for 16-QAM modulation with maximum 15 iterations compared to the MPA-based detector with maximum 30 iterations. It can be seen that with the over-relaxed iterative detection (labeled as OTFS-SOR-MRC with Init. Est. ($\omega=1.25$)), the BER performance is improved by around 2.5 dB at BER $=10^{-3}$. Moreover, the SOR-iterative algorithm converges on average in less than 8 iterations for SNR>15 dB. We can see from Fig. \ref{damp64Q} and \ref{ber_16qam_SOR} that the SOR parameter has more impact at higher modulation schemes, where  the initial low complexity estimate is less accurate and the convergence is generally slow without SOR.
     \begin{figure}
    \centering
    {\includegraphics[trim=10 0 0 10,clip,height=2.2in,width=3.4in]{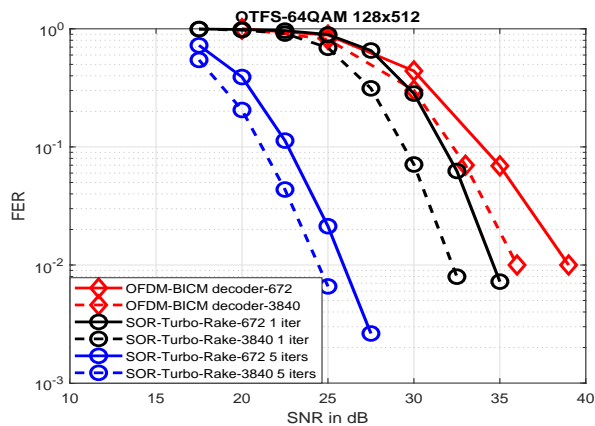}
     \vspace{-3mm}\caption{Turbo 64-QAM FER Plot: MRC vs BIC--MMSE-OFDM for codeword lengths: 672, 3840.}
    \label{comp_code}}
    \end{figure}
Fig. \ref{turboplot} and \ref{fer_64qam} shows the frame error performance of the plain and SOR-turbo-Rake decoder with initial low complexity estimate for 16 and 64 QAM modulation, respectively, compared with bit interleaved coded OFDM with MMSE detection scheme (labeled as OFDM BICM decoder). A half-rate LDPC code of length $N_c=3840$ bits from \cite{ldpc} is used 
 and every OTFS frame contains $\floor{NM\log_2(|\mathcal{Q}|)/N_c}$ codewords.

Turbo iterations are stopped when all the decoded codewords within the frame satisfy the LDPC parity check. It can be observed that just 1 iteration of turbo MRC detector (labeled as Turbo-Rake 1 iter) is required to achieve better error performance than the bit interleaved coded MMSE OFDM. 
Moreover, with the over-relaxation parameter $\omega=1.25$ (labeled as SOR-Turbo-Rake), a gain of $\approx 0.2$dB (for 16 QAM with 3 turbo iterations) and $\approx 1$dB (for 64 QAM with 3 turbo iterations) is achieved in the FER performance. The overall detector complexity in terms average number of iterations to converge is significantly reduced by using turbo iterations along with the initial estimates from the time-frequency single tap equalizer. 

Fig. \ref{comp_code} shows the FER performance of the proposed detector vs BICM-OFDM for different codeword lengths: long (labeled as SOR-Turbo-Rake-3840) and short (labeled as SOR-Turbo-Rake-672). For a fair comparison with the OFDM scheme, the FER plot for a single turbo iteration is also plotted alongside. It can be observed that, the proposed detector with single turbo iteration has a gain of $\approx 3$dB and $\approx 4$dB for codeword length of 3840 and 672, respectively, as compared to the OFDM scheme at a FER of $10^{-2}$. It can be noted that more iterations are required for short codewords to achieve the same performance as long codewords.

\section{Detector Complexity}
In the table below, we summarize and compare the overall complexity of the iterative Rake receiver (in terms of complex multiplications), including initial computations and Fourier domain transformations as discussed in Section \ref{sec1}.
 \begin{center}
\begin{tabular}{ | m{3.0cm} | m{0.6cm} | m{3.6cm} | } 
\hline
Computations per iteration & (I) & {$NM'(2L+1+2\log_2(N))$}  \\ 
\hline
\multirow{2}{3cm}{Initial computations} &  (II) & {$NM'(P+2L)$}  \\ 

\cline{2-3} &  (III) & {$NM[L+2\log_2(M)+3]$}   \\ 
\hline
\end{tabular}
\end{center}

Term (I) accounts for the computations inside each detector iteration, which includes calculating ${\Delta }\thicktilde{\bf g}_m^{(i)}$, $\Delta\thicktilde{\bf y}_{m}^{(i)}$, $\thicktilde{\bf c}_m^{(i)}$, and the symbol-vector hard decision estimates $\thicktilde{\bf x}_m^{(i)}$ in Algorithm \ref{algo2}. Term (II) is for initial computations, which involves calculating $M'L$ delay-time Doppler spread vectors $\thicktilde{\pmb \nu}_{m,l}$, initial $M'$ residual vectors $\Delta\thicktilde{\bf  y}_{m}^{(0)}$ in (\ref{deltaytd}), and $M'$ vectors $\thicktilde{\bf d}_m$ and term (III) is to compute the low complexity initial time-frequency estimate ${\hat {\bf x}}_m^{(0)}$ in (\ref{20}).

The detectors for OTFS with complexity linear in $NM$ and with non-ideal pulse shaping waveform (rectangular) are discussed in \cite{Ravi,kgp}. The complexity of the MPA detector per iteration scales with the number of paths on the discrete delay-Doppler grid and the alphabet size $|\mathcal{Q}|$, and has a complexity of the order $O(P|\mathcal{Q}|NM)$ \cite{Ravi}. The linear minimum mean square error detector proposed in \cite{kgp} even though is a non-iterative detector has a computational complexity of $O\left((l_{\max}^2+k_{\max}P^2)NM\right)$ whereas the proposed detector has a complexity of $O(SLNM)$ where $L \leq P$ and $S$ is the number of MRC detector iterations as given in Fig. \ref{ber_16qam_SOR}. 

\begin{figure}
    \centering
    {\includegraphics[trim=10 0 0 10,clip,height=2.3in,width=3.4in]{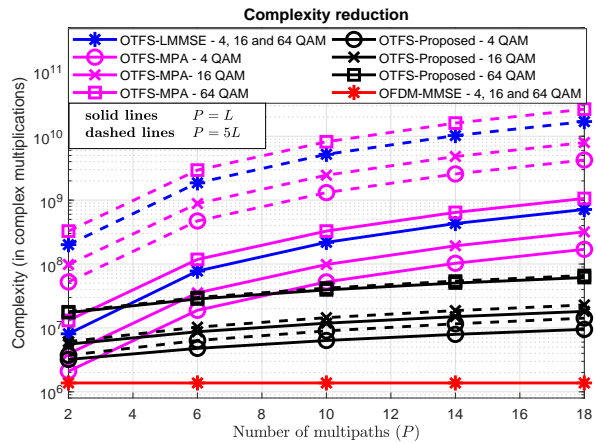}
     \vspace{-3mm}\caption{Complexity comparison with other linear detectors, for different modulation sizes, for an OTFS frame of size $N=128, M=512$ for $P=L$, i.e., for {\em one} Doppler path per delay tap (solid lines) and $P=5L$, i.e., for {\em five} Doppler paths per delay tap (dashed lines).}
    \label{comp}}
    \end{figure}

  The complexity of the proposed detector is compared with other linear complexity OTFS detectors, for different modulation sizes, number of multipaths in Fig. \ref{comp}. The dashed lines represents the case when there are 5 paths with distinct Doppler shifts in each delay tap i.e., $P=5L$. It can be concluded from Fig. \ref{comp} that the proposed detector complexity is significantly lower than the one of other OTFS detectors and closer to that of an OFDM single tap MMSE equalizer.   

For the iterative operation, the storage requirement for the MRC detector is $(L+2)NM$ complex numbers as only the $LNM$ delay-time channel coefficients, the $M$ RNPI vectors, and the $M'$ symbol vector estimates need to be stored for each iteration. For MPA, the storage requirement is much higher and of the order $O(P|\mathcal{Q}|NM)$, \cite{Ravi}.

\section{Conclusion}
We reformulated the OTFS input-output relation and proposed {\em two} versions of a {\em linear complexity} iterative rake detector algorithm for ZP-OTFS modulation based on the maximal ratio combining principle. We show that the MRC detector along with a low complexity initial estimate of symbol-vectors can achieve similar or better BER performance than the MPA detector with lower complexity and storage requirements. Based on the well studied Gauss-Seidel method, we introduced a successive over relaxation parameter to improve error performance and faster convergence of the proposed detector. The MRC detector performance was further improved with the aid of an outer error control coding scheme using turbo iterations. An additional advantage of the MRC detector is that the complexity is linear in $L$ (number of delay taps) rather than $P$ (total number of paths), thanks to the vector decomposition of the 2-D convolution with the channel.
\section*{Appendix}\label{appen}
\subsubsection{Proof of Lemma (\ref{rho})}
Consider the $M$ dimensional linear system of equations ${\bf z}_n={\bf R}_n\cdot{\bf s}_n$ without the noise term in (\ref{MRC_matrix}). The positive definite Hermitian system matrix ${\bf R}_n$ can be split as ${\bf D}_n+{\bf L}_n+{{\bf L}}^{\dag}_n$, where ${\bf D}_n$ and ${\bf L}_n \in \mathbb{C}^{M \times M}$ are the matrices containing the diagonal and strictly lower-triangular elements, respectively. Pre and post-multiplying both sides of (\ref{MRC_matrix}) by ${\bf D}_n^{-1/2}$ and ${\bf D}_n^{1/2}$, respectively, we get the re-scaled system of equations 
\begin{equation}
{\bf z}_n'={\bf R}_n'\cdot{\bf s}_n'
     \label{scale}
\end{equation}
where 
\begin{equation}
    {\bf R}_n'={\bf D}_n^{-1/2}\cdot{\bf R}_n\cdot{\bf D}_n^{-1/2},\quad{\bf z}_n'={\bf D}_n^{-1/2}\cdot{\bf z}_n,\quad{\bf s}_n'={\bf D}_n^{1/2}\cdot{\bf s}_n 
    \label{scalepar}
\end{equation}
${\bf R}_n^{\prime}$ is the re-scaled system matrix, which can be split as 
\begin{equation}
    {\bf R}^{\prime}_n={\bf I}_M+{\bf L}^{\prime}_n+\textbf{{\bf L}}^{\prime \dag}_n    
\end{equation}
where ${\bf L}_n^{\prime}={\bf D}_n^{-1/2}\cdot{\bf L}_n\cdot{\bf D}_n^{-1/2}$.

Since ${\bf R}_n^{\prime}$ is a positive definite Hermitian matrix, any non-zero vector ${\bf u}$ such that ${\bf u}^{\dag}\cdot{\bf u}=\beta>0$ satisfies,   
\begin{align}
    &{\bf u}^{\dag}\cdot({\bf I}_M+{\bf L}^{\prime}_n+{\bf L}^{\prime \dag}_n)\cdot{\bf u}>0 \nonumber
\\    \implies &\beta+2\Re{[{\bf u}^{\dag}\cdot{\bf L}^{\prime}_n\cdot{\bf u}]}>0.
    \label{posdef2}
\end{align}
The inequality in (\ref{posdef2}) can now be written as
\begin{equation}
a=\Re[{\bf u}^{\dag}\cdot{\bf L}^{\prime}_n\cdot{\bf u}]=\Re[{\bf u}^{\dag}\cdot{\bf L}^{\prime \dag}_n\cdot{\bf u}]>-\frac{\beta}{2} \label{ineq}
\end{equation}
where $\Re[\cdot]$ denotes the real part. Also note that 
\begin{equation}
b=\Im[{\bf u}^{\dag}\cdot{\bf L}^{\prime}_n\cdot{\bf u}]=-\Im[{\bf u}^{\dag}\cdot{\bf L}^{\prime \dag}_n\cdot{\bf u}] \label{ima}
\end{equation}
where $\Im[\cdot]$ denotes the imaginary part. 

Solving (\ref{MRC_matrix}) is equivalent to solving the linear system of equations in (\ref{scale}) and re-scaling its solution vector as given in (\ref{scalepar}). The equivalent GS iteration matrix ${\bf T}^{\rm GS}_n$ for (\ref{scalepar}) can be written as
\begin{align}
    {\bf T}^{\rm GS}_n=({\bf I}_M+{\bf L}^{\prime}_n)^{-1}\cdot{\bf L}^{\prime \dag}_n. \label{Tn3} 
\end{align} 
Now, the GS method for the system equation given in (\ref{Tn}) is guaranteed to converge if $|\lambda({\bf T}^{\rm GS}_n)|<1$, where $\lambda({\bf T}^{\rm GS}_n)$ denotes any eigenvalue of ${\bf T}^{\rm GS}_n$, which satisfy ${\bf T}^{\rm GS}_n\cdot {\bf v}=\lambda({\bf T}^{\rm GS}_n) {\bf v}$, for the corresponding eigenvectors ${\bf v}$, i.e.,
\begin{align}
    ({\bf I}_M+{\bf L}^{\prime}_n)^{-1}\cdot{\bf L}^{\prime \dag}_n\cdot {\bf v}=\lambda({\bf T}^{\rm GS}_n) {\bf v}  \label{Tn4}
.\end{align}
After multiplying both sides of (\ref{Tn4}) by ${\bf v}^{ H}\cdot({\bf I}_M+{\bf L}_n^{\prime})$, we can write $\lambda({\bf T}^{\rm GS}_n)$ as
\begin{align}
    \lambda({\bf T}^{\rm GS}_n)&=\frac{{\bf v}_n^{\dag}\cdot{\bf L}^{\prime \dag}_n\cdot {\bf v}_n}{ \beta+{\bf v}_n^{\dag}\cdot{\bf L}_n^{\prime}\cdot{\bf v}_n} =\frac{|a-j b|}{| \beta+a+j b|}
    &=\frac{\sqrt{a^2+b^2}}{\sqrt{(\beta+a)^2+b^2}}. \label{ineq2}
\end{align}
From (\ref{ineq}), (\ref{ima})  and (\ref{ineq2}), it can be seen that $|\lambda({\bf T}^{\rm GS}_n)|<1$.  Similarly for the case when ${\bf R}_n$ is positive semi-definite ,i.e., (\ref{ineq}) becomes $a \geq -\beta/2$, the eigenvalue inequality becomes $|\lambda({\bf T}^{\rm GS}_n)| \leq 1$. Since $\rho({\bf T}^{\rm GS}_n)$ is equal to the largest absolute value of the eigenvalues of ${\bf T}^{\rm GS}_n$, the positive definiteness of ${\bf R}_n$ ensures that $\rho({\bf T}^{\rm GS}_n)<1$.
\subsubsection{Proof of Lemma (\ref{omega_tn})}
Following the steps above, (\ref{ineq2}) can be modified for the eigenvalues of the  SOR-GS iteration matrix ${\bf T}_n^{\omega}$ defined in (\ref{new_tn}) as
\begin{align}
    \lambda({\bf T}_n^{\omega})=\frac{(\omega-1)({\bf v}^{\dag}\cdot{\bf v})+\omega({\bf v}^{\dag}\cdot{\bf L}^{\prime \dag}_n\cdot {\bf v}_n)}{ {\bf v}^{\dag}\cdot{\bf v}+\omega({\bf v}^{\dag}\cdot{\bf L}_n^{\prime}\cdot{\bf v}) }. \label{eig_omega}
\end{align}
 The condition for eigenvalues $\lambda({\bf T}^{\rm GS}_n)$ in (\ref{ineq2}) can then be modified for the SOR case as
\begin{align}
    |\lambda({\bf T}_n^{\omega})|
    &=\frac{\sqrt{((\omega-1)\beta+\omega a)^2+(\omega b)^2}}{\sqrt{(\beta+\omega a)^2+(\omega b)^2}}. \label{ineq3}
\end{align}
It can be seen from (\ref{ineq3}) that $|\lambda({\bf T}_n^{\omega})|<1$, if
$|(\omega-1)\beta+\omega a|<|\beta+\omega a|$, which is guaranteed if $0<\omega<2$.


\vspace{-2mm}
\begin{thebibliography}{1}
\bibitem{Hadani}
R. Hadani, S. Rakib, M. Tsatsanis, A. Monk, A. J. Goldsmith, A. F. Molisch, and R. Calderbank, ``Orthogonal time frequency space modulation,'' in {\it Proc. IEEE  Wireless Commun. Netw. Conf. (WCNC)}, San Francisco, CA, USA, Mar. 2017.
\bibitem{effDiv}
P. Raviteja, Y. Hong, E. Viterbo, and E. Biglieri, ``Effective Diversity of OTFS Modulation,'' {\em IEEE Wireless Commun. Lett.}, vol. 9, no. 2, pp. 249-253, Feb. 2020.
\bibitem{WCNC_paper}
T. Thaj and E. Viterbo, ``Low Complexity Iterative Rake Detector for Orthogonal Time Frequency Space Modulation,'' in {\it Proc. IEEE  Wireless Commun. Netw. Conf. (WCNC)}, Seoul, Korea (South), May 2020.
\bibitem{MRC0}
D. G. Brennan, ``Linear diversity combining techniques,'' in {\em Proc. IRE}, vol.47, pp. 1075-1102, June 1959.
\bibitem{MRC01}
S. Kondo and B. Milstein, ``Performance of multicarrier DS CDMA systems,'' {\em IEEE Trans. Commun.}, vol. 44, no. 2, pp. 238-246, Feb. 1996.
\bibitem{MRC1}
S. Imada and T. Ohtsuki, ``Pre-RAKE diversity combining for UWB systems in IEEE 802.15 UWB multipath channel,'' {\em Int. Workshop on UWBST and IWUWBS}, Kyoto, Japan, 2004, pp. 236-240.
\bibitem{MRC2}
Xiaofei Dong and N. C. Beaulieu, ``Optimal maximal ratio combining with correlated diversity branches,'' {\em IEEE Commun. Lett.,} vol. 6, no. 1, pp. 22-24, Jan. 2002.
\bibitem{GS1}
R. M. Buehrer, S. P. Nicoloso, and S. Gollamudi, ``Linear versus non-linear interference cancellation,''  {\em J. Commun. Netw.,} vol. 1, no. 2, pp. 118-133, Jun. 1999.

\bibitem{Ravi2}
P. Raviteja, Y. Hong, E. Viterbo, and E. Biglieri, ``Practical pulse-shaping waveforms for reduced-cyclic-prefix OTFS,'' {\em IEEE Trans. Veh. Technol.}, Oct. 2018.
\bibitem{Ravi3}
P. Raviteja, K. T. Phan, and Y. Hong, ``Embedded Pilot-Aided Channel Estimation for OTFS in Delay–Doppler Channels,'' {\em IEEE Trans. Veh. Technol.,} vol. 68, no. 5, pp. 4906-4917, May 2019.

\bibitem{D-OSDM}
T. Ebihara and G. Leus, ``Doppler-Resilient Orthogonal Signal-Division Multiplexing for Underwater Acoustic Communication,'' {\em IEEE J. Ocean. Eng.}, vol. 41, no. 2, pp. 408-427, April 2016.
\bibitem{UWA-channel}
M. Stojanovic and J. Preisig, ``Underwater acoustic communication channels: Propagation models and statistical characterization,'' {\em IEEE Commun. Mag.}, vol. 47, no. 1, pp. 84-89, Jan. 2009.
\bibitem{EVA}
``LTE Evolved Universal Terrestrial Radio Access (E-UTRA); Base Station (BS) radio transmission and reception,'' 3GPP TS 36.104 version 8.6.0 Release 8, Jul. 2009, ETSI TS.
doi: 10.1109/MCOM.2009.4752682
\bibitem{Wireless_book}
D. Tse and P. Viswanath, {\em Fundamentals of Wireless Communication}, 3{rd} ed. Cambridge University Press, ISBN: 9780521845274, 2005.  
\bibitem{farhang}
A. Farhang, A. RezazadehReyhani, L. E. Doyle and B. Farhang-Boroujeny, ``Low complexity modem structure for OFDM-based orthogonal time frequency space modulation,'' in {\em IEEE Wireless Commun. Lett.,} vol. 7, no. 3, pp. 344-347, Jun. 2018.
\bibitem{Hadani2}
R. Hadani, S. Rakib, M. Tsatsanis, A. Monk, A. J. Goldsmith, A. F. Molisch, and R. Calderbank, ``Orthogonal time frequency space modulation,'' Aug. 2018.
doi: arXiv:1808.00519
\bibitem{perm}
C. F. V. Loan, ``The ubiquitous Kronecker product,'' {\em J. Comput. Appl. Math.,} vol. 123, issues 1–2, pp. 85-100, ISSN 0377-0427, 2000.
\bibitem{sim_mat}
R. A. Beezer, {\em A First Course in Linear Algebra}, 3{rd} ed. Washington, DC, USA, Congruent Press, pp. 404-406, ISBN: 978-0-9844175-5-1, 1973.
\bibitem{LSBook}
A. Björck, {\em Numerical Methods for Least Squares Problems,} SIAM, 1996. doi: 10.1137/1.9781611971484
\bibitem{GSBook}
Y. Saad, {\em Iterative Methods for Sparse Linear Systems}, 2{nd} ed. SIAM, 2003. doi: 10.1137/1.9780898718003
\bibitem{mathJourn}
A. Hadjidimos, ``Successive overrelaxation (SOR) and related methods,'' {\em J. Comput. Appl. Math.,} vol. 123, Issues 1–2, pp. 177-199, ISSN 0377-0427, Nov. 2000. 
\bibitem{Ravi}
P.\ Raviteja, K.T. Phan, Yi Hong, and E. Viterbo, ``Low-complexity iterative detection for orthogonal time frequency space modulation,'' in {\em Proc. IEEE Wireless Commun. Netw. Conf. (WCNC),} Barcelona, Apr. 2018.
\bibitem{Ravi1}
P.\ Raviteja, K.T. Phan, Yi Hong, and E. Viterbo, ``Interference cancellation and iterative detection for orthogonal time frequency space modulation,'' {\em IEEE Trans. Wireless Commun.}, vol. 17, no. 10, pp. 6501-6515, Oct. 2018.


\bibitem{kgp}
S. Tiwari, S.S. Das and V. Rangamgari, ``Low-complexity LMMSE receiver for OTFS,'' {\em IEEE Commun. Lett.}, Oct. 2019.

\bibitem{ldpc}
T.T.B.\ Nguygen, T.N.\ Tan and H.\ Lee, ``Effecient QC-LDPC Encoder for 5G New Radio,'' {\em Electronics}, vol. 8, no. 6, p. 668, Jun. 2019. 

\end{thebibliography}
\end{document}